\documentclass[twocolumn]{aastex7}
\shorttitle{Electromagnetic Counterpart Candidates for GW231123}
\shortauthors{L. He et al.}
\graphicspath{{./}}
\usepackage{times}
\usepackage{amsmath}
\usepackage{hyperref}
\begin{document}
\title{Searching for Electromagnetic Counterpart Candidates to GW231123}

\correspondingauthor{Lei He, Zhengyan Liu, Wen Zhao}
\author[0000-0001-7613-5815]{Lei He}
\affiliation{Department of Astronomy, University of Science and Technology of China, Hefei, Anhui 230026, People’s Republic of China}
\affiliation{School of Astronomy and Space Sciences, University of Science and Technology of China, Hefei, Anhui 230026, People’s Republic of China}
\email[show]{helei0831@mail.ustc.edu.cn}

\author[0000-0001-7688-6504]{Liang-Gui Zhu}
\affiliation{Department of Astronomy, University of Science and Technology of China, Hefei, Anhui 230026, People’s Republic of China}
\affiliation{School of Astronomy and Space Sciences, University of Science and Technology of China, Hefei, Anhui 230026, People’s Republic of China}
\email{lianggui.zhu@ustc.edu.cn}

\author[0000-0002-2242-1514]{Zheng-Yan Liu}
\affiliation{Department of Astronomy, University of Science and Technology of China, Hefei, Anhui 230026, People’s Republic of China}
\affiliation{School of Astronomy and Space Sciences, University of Science and Technology of China, Hefei, Anhui 230026, People’s Republic of China}
\email[show]{ustclzy@mail.ustc.edu.cn}

\author[0000-0001-9098-6800]{Rui Niu}
\affiliation{Department of Astronomy, University of Science and Technology of China, Hefei, Anhui 230026, People’s Republic of China}
\affiliation{School of Astronomy and Space Sciences, University of Science and Technology of China, Hefei, Anhui 230026, People’s Republic of China}
\email{nrui@ustc.edu.cn}

\author[0009-0004-8038-4741]{Chao Wei}
\affiliation{Department of Astronomy, University of Science and Technology of China, Hefei, Anhui 230026, People’s Republic of China}
\affiliation{School of Astronomy and Space Sciences, University of Science and Technology of China, Hefei, Anhui 230026, People’s Republic of China}
\email{charles272@mail.ustc.edu.cn}

\author[0009-0002-0900-9901]{Bing-Zhou Gao}
\affiliation{Department of Astronomy, University of Science and Technology of China, Hefei, Anhui 230026, People’s Republic of China}
\affiliation{School of Astronomy and Space Sciences, University of Science and Technology of China, Hefei, Anhui 230026, People’s Republic of China}
\email{supernova_gbz@mail.ustc.edu.cn}

\author[0009-0002-5634-8842]{Ming-Shen Zhou}
\affiliation{Department of Astronomy, University of Science and Technology of China, Hefei, Anhui 230026, People’s Republic of China}
\affiliation{School of Astronomy and Space Sciences, University of Science and Technology of China, Hefei, Anhui 230026, People’s Republic of China}
\email{zmetassin@mail.ustc.edu.cn}

\author[0000-0001-6223-840X]{Run-Duo Liang}
\affiliation{National Astronomical Observatories, Chinese Academy of Sciences, 20A Datun Road, Beijing 100101, People’s Republic of China}
\email{liangrd@bao.ac.cn}

\author[0000-0001-8955-0452]{Ken Chen}
\affiliation{Department of Astronomy, University of Science and Technology of China, Hefei, Anhui 230026, People’s Republic of China}
\affiliation{School of Astronomy and Space Sciences, University of Science and Technology of China, Hefei, Anhui 230026, People’s Republic of China}
\email{cken@ustc.edu.cn}

\author[0000-0001-9449-9268]{Jian-Min Wang}
\affiliation{Key Laboratory for Particle Astrophysics, Institute of High Energy Physics, Chinese Academy of Sciences, 19B Yuquan Road, Beijing 100049, People’s Republic of China}
\email{wangjm@ihep.ac.cn}

\author[0000-0002-7152-3621]{Ning Jiang}
\affiliation{Department of Astronomy, University of Science and Technology of China, Hefei, Anhui 230026, People’s Republic of China}
\affiliation{School of Astronomy and Space Sciences, University of Science and Technology of China, Hefei, Anhui 230026, People’s Republic of China}
\email{jnac@ustc.edu.cn}

\author[0000-0002-4223-2198]{Zhen-Yi Cai}
\affiliation{Department of Astronomy, University of Science and Technology of China, Hefei, Anhui 230026, People’s Republic of China}
\affiliation{School of Astronomy and Space Sciences, University of Science and Technology of China, Hefei, Anhui 230026, People’s Republic of China}
\email{zcai@ustc.edu.cn}

\author[0000-0002-9092-0593]{Ji-an Jiang}
\affiliation{Department of Astronomy, University of Science and Technology of China, Hefei, Anhui 230026, People’s Republic of China}
\affiliation{School of Astronomy and Space Sciences, University of Science and Technology of China, Hefei, Anhui 230026, People’s Republic of China}
\email{jian.jiang@ustc.edu.cn}

\author[0000-0002-7835-8585]{Zi-Gao Dai}
\affiliation{Department of Astronomy, University of Science and Technology of China, Hefei, Anhui 230026, People’s Republic of China}
\affiliation{School of Astronomy and Space Sciences, University of Science and Technology of China, Hefei, Anhui 230026, People’s Republic of China}
\email{daizg@ustc.edu.cn}

\author[0000-0002-7330-4756]{Ye-Fei Yuan}
\affiliation{Department of Astronomy, University of Science and Technology of China, Hefei, Anhui 230026, People’s Republic of China}
\affiliation{School of Astronomy and Space Sciences, University of Science and Technology of China, Hefei, Anhui 230026, People’s Republic of China}
\email{yfyuan@ustc.edu.cn}

\author[0000-0003-1720-9727]{Jian Li}
\affiliation{Department of Astronomy, University of Science and Technology of China, Hefei, Anhui 230026, People’s Republic of China}
\affiliation{School of Astronomy and Space Sciences, University of Science and Technology of China, Hefei, Anhui 230026, People’s Republic of China}
\email{jianli@ustc.edu.cn}

\author[0000-0001-8738-6011]{Saurabh W. Jha}
\affiliation{Departmentss of Physics and Astronomy, Rutgers, the State University of New Jersey, Piscataway, NJ 08854, USA}
\email{saurabh@physics.rutgers.edu}

\author[0000-0002-1330-2329]{Wen Zhao}
\affiliation{Department of Astronomy, University of Science and Technology of China, Hefei, Anhui 230026, People’s Republic of China}
\affiliation{School of Astronomy and Space Sciences, University of Science and Technology of China, Hefei, Anhui 230026, People’s Republic of China}
\affiliation{College of Physics, Guizhou University, Guiyang, 550025, People’s Republic of China}
\email[show]{\\ wzhao7@ustc.edu.cn}

\begin{abstract}
  The detection of GW231123, a gravitational-wave (GW) event with exceptionally massive and rapidly spinning black holes, suggests the possible formation within an active galactic nucleus (AGN) disk, which provides a favorable environment for potentially generating an observable electromagnetic (EM) counterpart. We conduct a search for such a counterpart by crossmatching the GW localization with a comprehensive catalog of AGN flares from the Zwicky Transient Facility. Our analysis yields six plausible optical flare candidates that are spatially and temporally coincident with GW231123 and exhibit significant deviations from their AGN baseline flux. These candidates provide a well-defined starting point for evaluating a possible EM counterpart of GW231123. To facilitate future verification, we present a set of observational diagnostics that can be used to confirm or rule out their association, enabling a more systematic exploration of the AGN formation channel in upcoming multi-messenger searches.
\end{abstract}

\keywords{\uat{Active galactic nuclei}{16}; \uat{Gravitational wave sources}{677}; \uat{Stellar mass black holes}{1611}}

\section{Introduction}

The recent discovery of GW231123 reported by the LIGO-Virgo-KAGRA (LVK) collaboration has introduced the most extreme binary black hole (BBH) system yet \citep{theligoscientificcollaborationGW231123BinaryBlack2025}. Its component masses, inferred to be $137^{+22}_{-17} M_{\odot}$ and $103^{+20}_{-52} M_{\odot}$, make it the most massive BBH merger detected so far. Moreover, the system is distinguished by the exceptionally high dimensionless spins of both components, $\chi_1= 0.9^{+0.10}_{-0.19}$ and $\chi_2 = 0.8^{+0.20}_{-0.51}$, which are the largest spin values ever reported from a gravitational-wave (GW) observation.

The unusually large component masses of GW231123 present a serious challenge to standard stellar evolution models, as both components fall within the predicted pair-instability mass gap \citep{fowlerNeutrinoProcessesPair1964, woosleyPulsationalPairinstabilitySupernovae2017, maganahernandezEvidenceNewFeature2025}. This suggests the need for alternative formation channels \citep{mapelliFormationChannelsSingle2021, theligoscientificcollaborationGW231123BinaryBlack2025}, with hierarchical mergers providing a particularly compelling explanation. In this scenario, black holes (BHs) are assembled through successive mergers, a process that naturally produces massive BHs capable of bridging the mass gap. Furthermore, the high component spins are also consistent with predictions that hierarchically merged BHs tend to cluster around a characteristic value of $\sim$0.7 \citep{gerosaHierarchicalMergersStellarmass2021}. Several studies have already explored this pathway for GW231123. \citet{liGW231123ProductSuccessive2025} suggest that its components originate from the successive mergers of $\sim$6 and $\sim$4 first-generation BHs, respectively. \citet{delfaveroProspectsFormationGW2311232025} propose that GW231123 is most consistent with a merger of fourth- and third-generation BHs. \citet{liHierarchicalMergerScenario2025} show that 2G+2G merger scenario for GW231123 is favored over a 2G+1G (or 3G+2G) merger within a Bayesian framework.

Beyond hierarchical mergers, several other scenarios have also been proposed for the origin of GW231123. These include isolated binary evolution of very massive and rapidly rotating helium cores \citep{croonCanStellarPhysics2025,gottliebSpinningGapDirectHorizon2025}, in some cases through chemically homogeneous evolution \citep{stegmannResolvingBlackHole2025,popaVeryMassiveRapidly2025}, as well as other possibilities such as sustained accretion \citep{bartosAccretionAllYou2025,kirogluHierarchicalMergersAccretiondriven2025}, strong gravitational lensing \citep{shanDegeneracyMicrolensingWave2025}, formation from Population III stars \citep{tanikawaGW231123FormationPopulation2025,liuFormationGW231123Population2025,wuPairinstabilityGapBlack2025}, and primordial BHs \citep{yuanGW231123MassGap2025,lucaGW231123PossiblePrimordial2025,nojiriDynamicalBlackHole2025}.

Regardless of the specific formation channels for GW231123, the AGN disk provides a particularly favorable environment for assembling such massive BHs \citep{McKernan2012,Tagawa2020}. The dense gas can promote both the formation and evolution of binaries \citep{Li2023,Dittmann2024}, while the deep potential well of the central supermassive black hole (SMBH) helps retain merger remnants that would otherwise be ejected \citep{varmaEvidenceLargeRecoil2022}, and gas friction tends to drag the inclined remnants back into the disk \citep{Fabj2020}. Crucially, BBH mergers embedded in AGN disks are expected to have electromagnetic (EM) counterparts \citep{bartosRapidBrightStellarmass2017}. In these scenarios, the EM emission is thought to arise from interactions between the binary (or its merger remnant) and the surrounding gas. Proposed mechanisms include ram pressure stripping of gas bound to the remnant \citep{mckernanRampressureStrippingKicked2019}, hyper-Eddington accretion capable of driving powerful outflow explosions \citep{wangAccretionmodifiedStarsAccretion2021,Kimura2021,Rodriguez-Ramirez2025}, and the launching of a jet driven by the accreting remnant \citep{tagawaShockCoolingBreakout2024,Rodriguez-Ramirez2024,chenElectromagneticCounterpartsPowered2024}--all of which produce more luminous emission with increasing BH mass, just as the case of GW231123.

Identifying an EM counterpart to a BBH merger would have profound scientific implications. It would not only shed light on the origin and evolution of stellar-mass black holes and stellar populations within AGN disks, but also provide valuable insight into the physical processes governing AGN environments \citep{vajpeyiMeasuringPropertiesActive2022}. Moreover, confirmed counterparts could serve as standard sirens, enabling independent cosmological measurements of the Hubble constant and other key parameters through combined GW and EM observations \citep{mukherjeeFirstMeasurementHubble2020,chenStandardSirenCosmological2022,bomStandardSirenCosmology2024b,liUseBinaryBlack2025}.

The search for EM counterparts to BBH mergers has yielded several candidates in previous observing runs \citep{grahamCandidateElectromagneticCounterpart2020,grahamLightDarkSearching2023,cabreraSearchingElectromagneticEmission2024a}, though confirming a definitive association remains a major challenge. Even for the most notable case, the flare ZTF19abanrhr associated with GW190521, conflicting interpretations persist in the literature \citep{ashtonCurrentObservationsAre2021,palmeseLIGOVirgoBlack2021,mortonGW190521BinaryBlack2023}. A recent reassessment by \citet{heTracingLightIdentification2025a} found that only two of seven previously proposed candidates remain statistically consistent with their associated GW events after three years.

Motivated by the possibility of a detectable EM counterpart from an AGN disk associated with GW231123, we carry out a systematic search for such an optical flare. Using a catalog of AGN flares constructed from Zwicky Transient Facility (ZTF) data \citep{heSystematicSearchAGN2025}, we perform a rigorous crossmatch to search for flares that are spatially and temporally coincident with GW231123, which yields six potential counterparts. The rest of this paper is organized as follows. In Section \ref{sec:data}, we describe the GW data and our search methodology. We present our results of search in Section \ref{sec:results} and discuss the possible interpretations in Section \ref{sec:discussion}. Finally, we draw the conclusions in Section~\ref{sec:conclusion}. Throughout this work we use a Planck 2018 cosmology with $H_0 = 67.7 \mathrm{km \ s^{-1} Mpc^{-1}}$, $\Omega_\Lambda=0.689$, and $\Omega_m = 0.311$ \citep{aghanimPlanck2018Results2020}.

\section{Data and Searching Method\label{sec:data}}
\subsection{GW data and AGN flares}
\begin{figure*}[htb]
  \centering
  \gridline{\fig{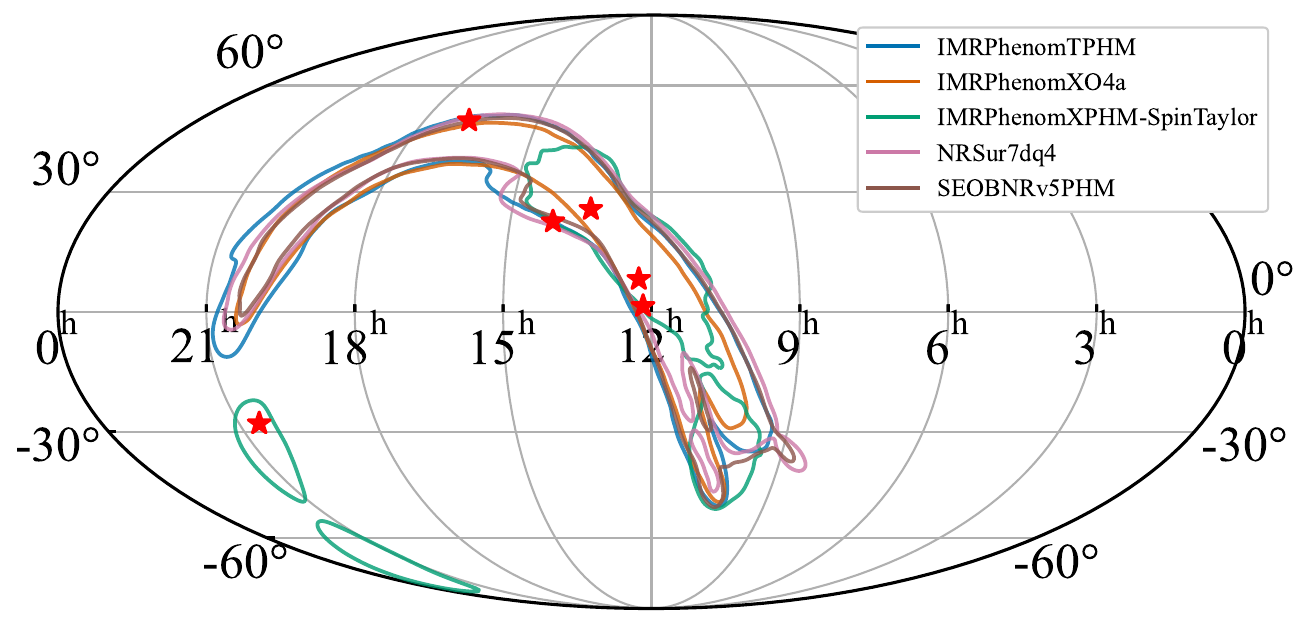}{\textwidth}{}
  }
  \caption{The $99\%$ credible localization regions from different waveforms. The red stars in left panel mark the positions of our final candidates. }
  \label{fig:GW-skymap}
\end{figure*}

\begin{figure}[htb]
  \fig{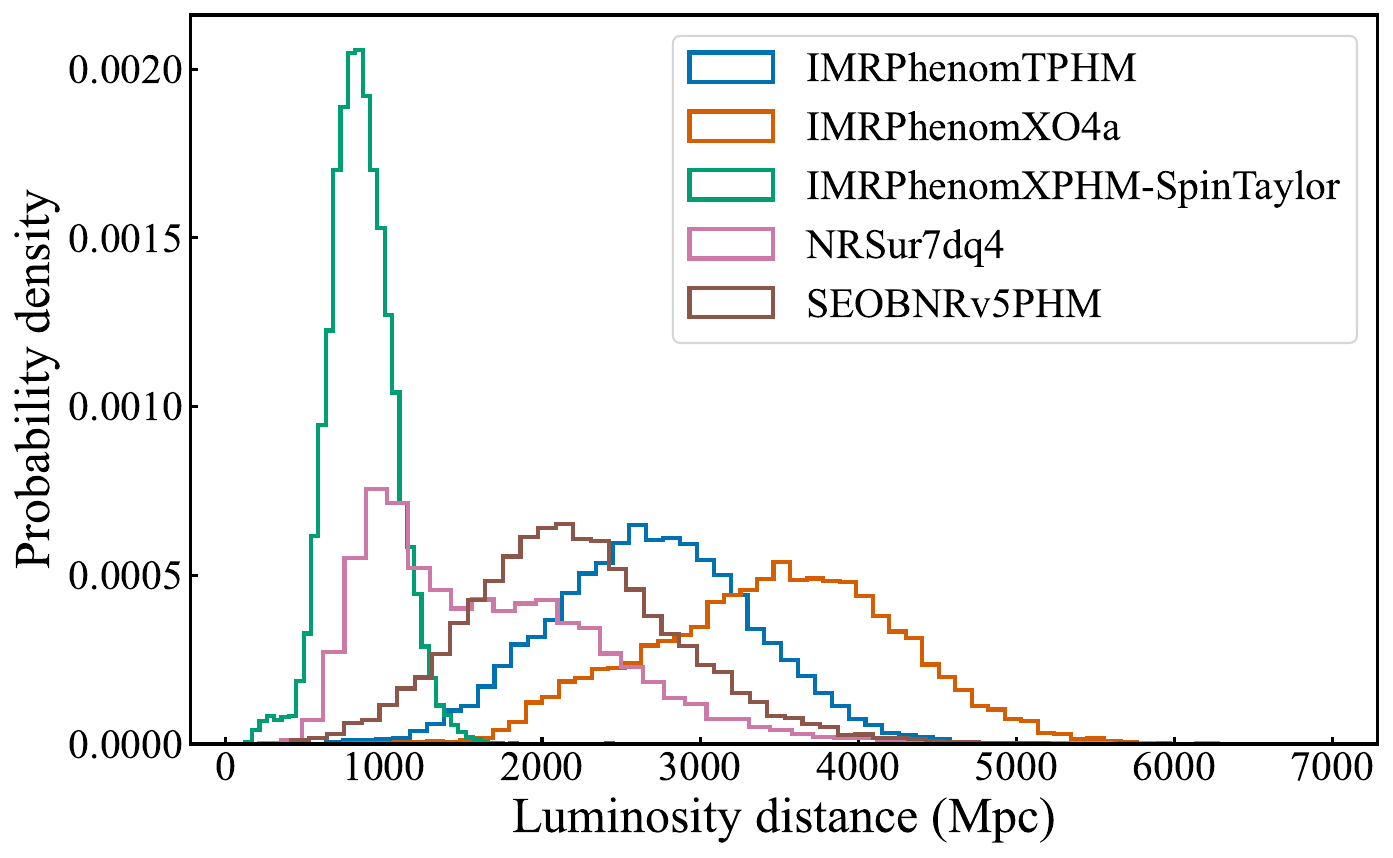}{0.5\textwidth}{}
  \caption{The luminosity distance posterior distributions from different waveforms.}
  \label{fig:dL}
\end{figure}

The event GW231123 was observed by both the Advanced LIGO Hanford and Livingston detectors on 2023 November 23 at 13:54:30 UTC, with a signal-to-noise ratio of approximately 20 \citep{theligoscientificcollaborationGW231123BinaryBlack2025}. Some of its key properties are presented in Table~\ref{tab:gwdata}. The $90\%$ confidence area spans about 400-1000 $\mathrm{deg}^2$, while the inferred luminosity distance is estimated to range from several hundred to several thousand Mpc, depending on the waveform model employed. The significant disagreement between models, illustrated in Figure~\ref{fig:GW-skymap} and Figure~\ref{fig:dL}, likely reflects systematic waveform uncertainties. In particular, all models show strong support for component spins exceeding 0.8, while current precessing signal models are not calibrated against numerical relativity simulations in this high-spin regime. This limitation may contribute to the observed differences in the inferred posteriors \citep{theligoscientificcollaborationGW231123BinaryBlack2025}.

\begin{deluxetable}{ll}[htb]
  \tablecaption{Source Properties of GW231123}
  \tablehead{\colhead{Property \hspace*{5cm}} & \colhead{Value}}
  \startdata
  Primary mass ($M_{\odot}$) & $137^{+22}_{-17}$ \\
  Secondary mass ($M_{\odot}$) & $103^{+20}_{-52}$ \\
  Mass ratio & $0.75^{+0.22}_{-0.39}$\\
  Total mass ($M_{\odot}$)& $238^{+28}_{-49}$ \\
  Final mass ($M_{\odot}$)& $225^{+26}_{-43}$ \\
  Primary spin magnitude & $0.90^{+0.10}_{-0.19}$\\
  Secondary spin magnitude & $0.80^{+0.20}_{-0.51}$\\
  Effective inspiral spin & $0.31^{+0.24}_{-0.39}$\\
  Effective precessing spin & $0.77^{+0.17}_{-0.19}$\\
  Final spin & $0.84^{+0.08}_{-0.16}$\\
  Kick velocity ($\mathrm{km \ s^{-1}}$)& $967^{+829}_{-857}$ \\
  \enddata
  \label{tab:gwdata}
\tablecomments{These are median values and 90\% credible intervals reported by \citet{theligoscientificcollaborationGW231123BinaryBlack2025}.}
\end{deluxetable}

To search for a potential counterpart to GW231123 among AGNs, we employ the AGN Flare Coarse Catalog (AGNFCC) from \citet{heSystematicSearchAGN2025}. This catalog, constructed from a systematic search for AGN flares based on ZTF Data Release 23, provides a comprehensive list of AGN flares up to 2024 October 31\footnote{\url{https://github.com/Lyle0831/AGN-Flares}}. It contains the source positions, redshifts (when available), and fitted peak times of the flares. For sources without redshifts, we supplement the catalog with photometric redshifts from Pan-STARRS1\citep{beckPS1STRMNeuralNetwork2021} when available.

\subsection{Searching method \label{sec:method}}

Given the uncertainties in current waveform models and the disagreements among them, we select all AGN flares located within the 99\% credible localization volume of GW231123 for each waveform model. The crossmatch is performed within the 3D credible volume for sources with a known redshift (spectroscopic or photometric) and within the 2D credible area for those without. This procedure is implemented using the \texttt{crossmatch} method from the \texttt{ligo.skymap} Python package\footnote{\url{https://lscsoft.docs.ligo.org/ligo.skymap}}. By further requiring that the fitted flare peak time occurred after the GW231123 trigger, we find 166 candidate AGN flares that could potentially be associated with GW231123.

For all selected AGN flares, we retrieve their historical g and r band light curves from the ZTF forced photometry service \citep{masciNewForcedPhotometry2023}, following the recommended procedures\footnote{\url{https://irsa.ipac.caltech.edu/data/ZTF/docs/ztf\_forced\_photometry.pdf}}. The data span from March 2018 to October 2025, extending the temporal baseline by nearly one year compared to \citet{heSystematicSearchAGN2025}. We then apply the flare detection algorithm described in \citet{heSystematicSearchAGN2025} to this extended dataset. Specifically, we first identify potential flares using the Bayesian Block algorithm. We then model the long-term AGN variability using a Gaussian process and infer the corresponding hyperparameters from the observed light curve. Using these hyperparameters, we generate 10,000 simulated light curves with the same cadence and photometric uncertainties as the data. For each simulation, we compute the test statistic $\lambda$ over the flare interval, where larger values of $\lambda$ correspond to stronger deviations from stochastic variability. We compare the observed $\lambda$ to the simulated distribution and quantify the rarity of the event as the fraction of simulations with $\lambda$ exceeding the observed value. Only events above the 99.5th percentile of the simulated distribution are retained, thereby minimizing contamination from intrinsic AGN variability.

Unlike the AGNFCC constructed using light curves based on single-exposure images, we analyze forced-photometry light curves extracted at fixed source positions. This approach reduces host galaxy contamination presented in single-exposure science images. Combined with the longer temporal baseline, it enables a more reliable characterization of long-term AGN variability, making the analysis more accurate and robust. Since our search focuses on flares occurring after the GW trigger, the inclusion of the most recent data is particularly important for identifying potential counterparts to GW231123.

Following the flare detection, we fit the light curve of each candidate with a Gaussian rise-exponential decay model, as adopted in \citet{grahamLightDarkSearching2023}:
\begin{equation}
  \label{eq:flare}
  y(t) = \left\{
    \begin{aligned}
      &r_{0} + A\exp\left(-\frac{(t-t_{0})^{2}}{2t_{\mathrm{g}}^{2}}\right), t \leq t_{0} \\
      &r_{0} + A\exp\left(-\frac{(t-t_{0})}{t_{\mathrm{e}}}\right), t \ge t_{0},
    \end{aligned}
    \right.
  \end{equation}where $r_0$ is the baseline flux level, $A$ is the flare amplitude, $t_0$ is the peak time of the flare, and $t_{\mathrm{g}}$ and $t_{\mathrm{e}}$ are the rise and decay timescales, respectively. This modeling, combined with visual inspection, allows us to apply a set of strict criteria to identify events that could be associated with GW231123. We first exclude flares flagged in the AGNFCC as known transients, such as supernovae, tidal disruption events, or blazars. For the remaining candidates, we require that the flare begins to rise only after the GW trigger and that it appears as a single, isolated outburst in the AGN's light curve. Finally, we retain only flares with rise and decay timescales between 3 and 200 days. This range effectively excludes short-lived artifacts, given the ZTF survey cadence, as well as long-term variability that is more likely to be attributable to intrinsic AGN activity.

  \subsection{Odds ratio for AGN-GW Association \label{sec:odds}}

To quantify the likelihood that an AGN flare is associated with the GW event rather than arising from a chance coincidence, we compute the odds ratio following the framework presented in \citet{heTracingLightIdentification2025a}.

In this framework, two competing hypotheses are considered: the association model ($H_A$), in which the AGN flare and the GW event originate from the same astrophysical source, and the coincidence model ($H_C$), in which the flare and the GW event are unrelated. The relative support for these two hypotheses is quantified through the odds ratio
\begin{equation}
O_C^A = B_C^A \times P_C^A,
\end{equation}
where $B_C^A$ denotes the Bayes factor and $P_C^A$ represents the prior odds between the association and coincidence models.

The Bayes factor is defined as the ratio of the evidences for the two models. The evidence under a given hypothesis $\mathcal{H}_i$ can be written as
\begin{equation}
  \begin{aligned}
  p(d|\mathcal{H}_i) = & \int \frac{p(M_1^{\mathrm{eff}}, D_L^{\mathrm{eff}},\alpha,\delta|d)}{p(D_L^{\mathrm{eff}})} \\
  &\times p(M_1,z,\alpha,\delta | \mathcal{H}_i) dM_1dzd\alpha d\delta,
  \end{aligned}
\end{equation}
\added{Here $M_1$ is the source-frame primary mass, $z$ is the cosmological redshift, and $\alpha$ and $\delta$ denote the right ascension and declination of the source. The quantities $M_1^{\mathrm{eff}}$ and $D_L^{\mathrm{eff}}$ are the effective detector-frame mass and luminosity distance, which are obtained from the source-frame parameters.}

In the coincidence model, the GW event and the AGN flare are assumed to be unrelated. We adopt uniform priors for sky position and a redshift prior proportional to the comoving volume. \added{The source-frame primary mass prior is taken to be the \textsc{Broken Power Law + 2 Peaks} model inferred from the GWTC-4 events \citep{collaborationGWTC40PopulationProperties2025}. The effective primary mass and luminosity distance are simply given by}
\begin{equation}
  M_1^{\mathrm{eff}} = M_1 (1+z), \quad D_L^{\mathrm{eff}} = D_L(z).
\end{equation}

In the association model, the GW event is assumed to originate from a specific AGN, so the sky position is fixed at $(\alpha_{\mathrm{AGN}}, \delta_{\mathrm{AGN}})$ and the redshift is set to $z_{\mathrm{AGN}}$. \added{For our fiducial calculation, the source-frame primary mass prior follows the AGN disk BBH mass distribution proposed by \citet{vaccaroImpactGasHardening2024}. When mapping the intrinsic source-frame parameters to the effective detector-frame quantities, we include the environmental effects of the AGN disk, such as relativistic and gravitational redshifts following the procedure described in \citet{mortonGW190521BinaryBlack2023}. The details are provided in Appendix~\ref{appendix:environment}.}

The prior odds represent the expected probability of an AGN flare being physically associated with the GW event before considering the observational data. Following \citet{heTracingLightIdentification2025a}, the prior odds are estimated based on the number of AGNs located within the GW localization volume that could potentially produce detectable flares. If $N$ such AGNs are present, the prior odds can be approximated as $P_C^A \approx 1/N$. In practice, we estimate the number of potential AGNs flares as $N \approx n_s (1 - p_{\mathrm{flare}})$, where $n_s$ is the number of AGNs located within the 99\% GW localization volume, computed using the AGN catalogs compiled in \citet{heSystematicSearchAGN2025}, and $1 - p_{\mathrm{flare}}$ represents the probability that an observed flare is consistent with stochastic AGN variability as estimated from the Gaussian process simulations.

\section{Results\label{sec:results}}
\begin{figure*}[htb!]
  \centering

  \gridline{
    \fig{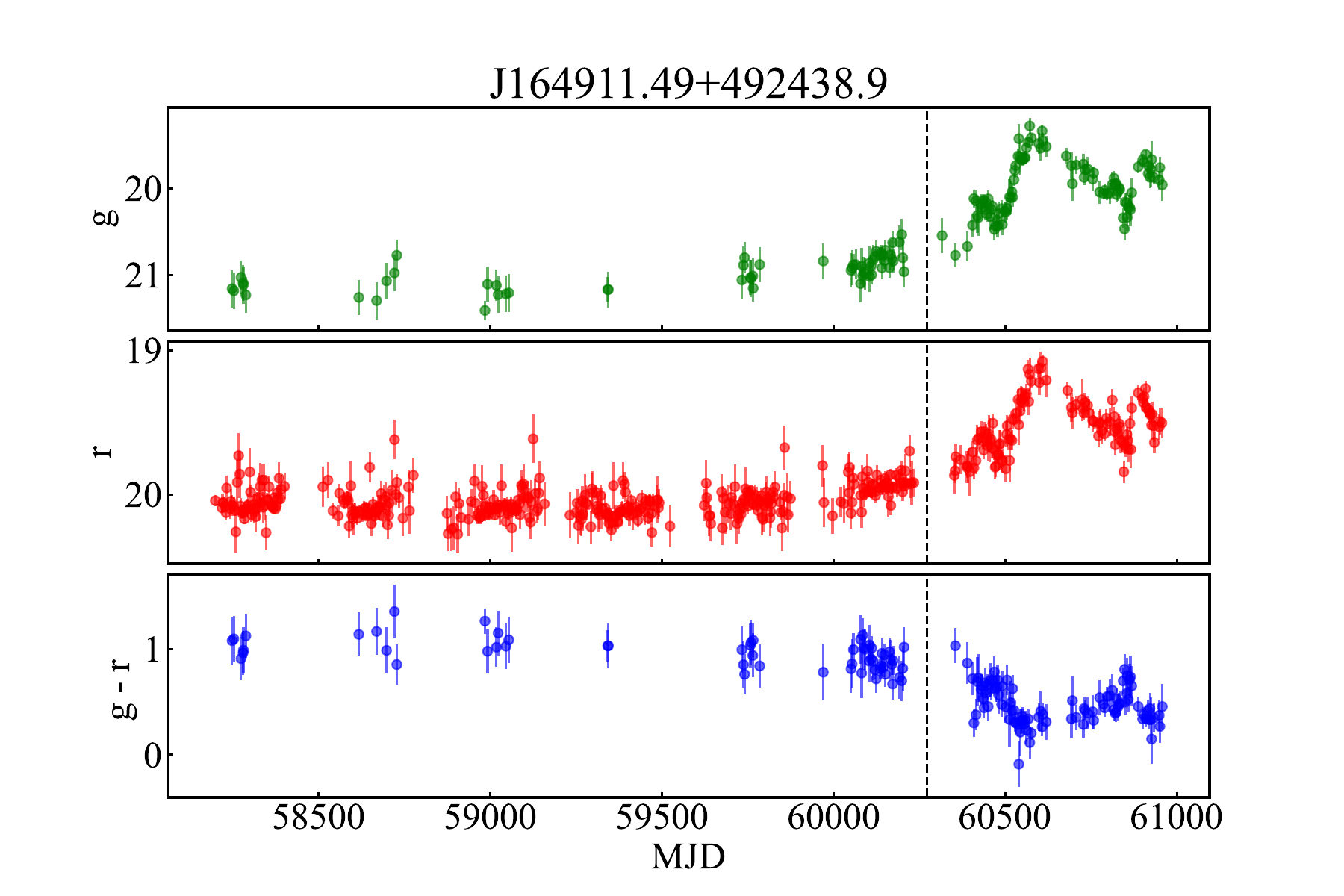}{0.5\textwidth}{}
    \fig{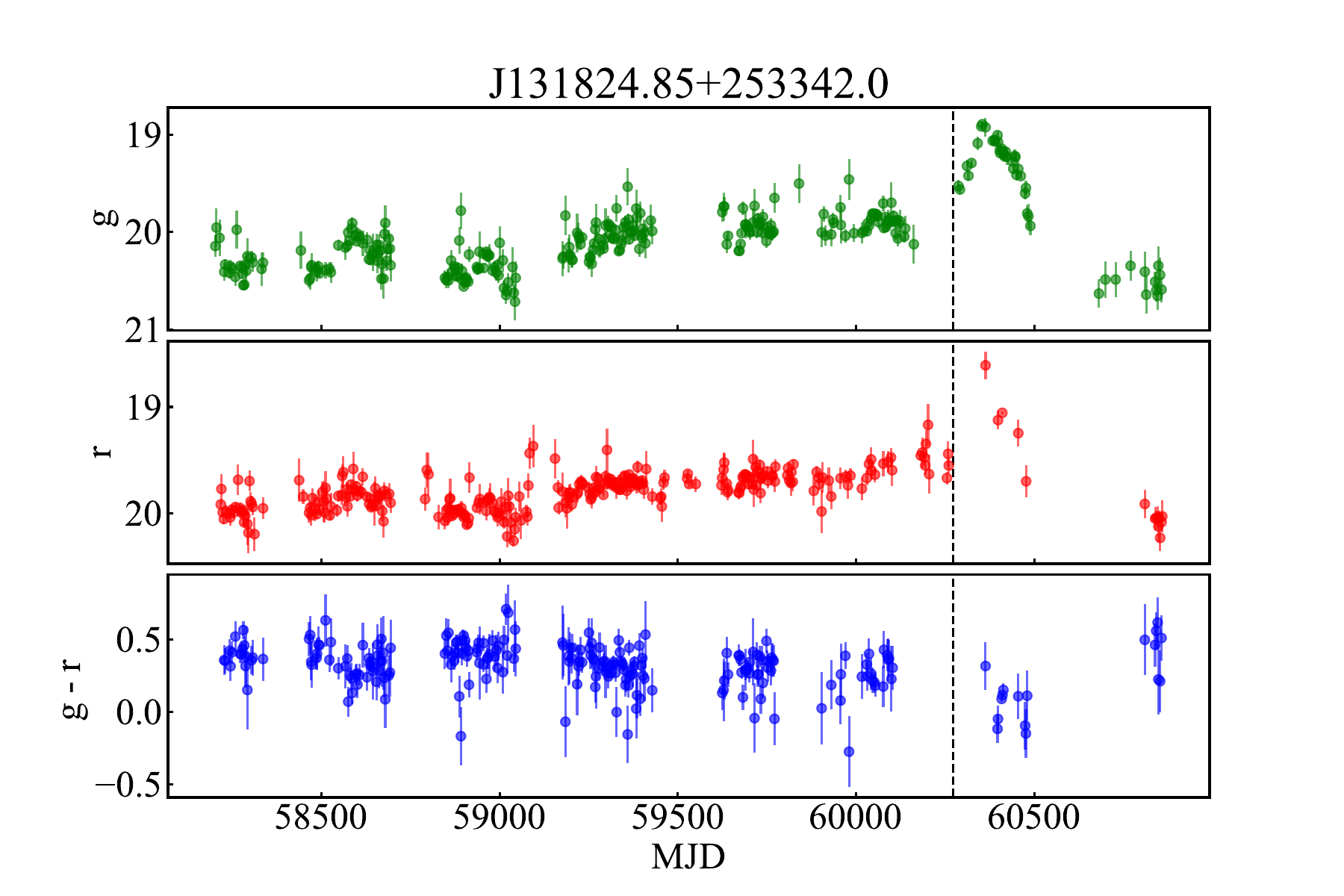}{0.5\textwidth}{}
  }
  \gridline{
    \fig{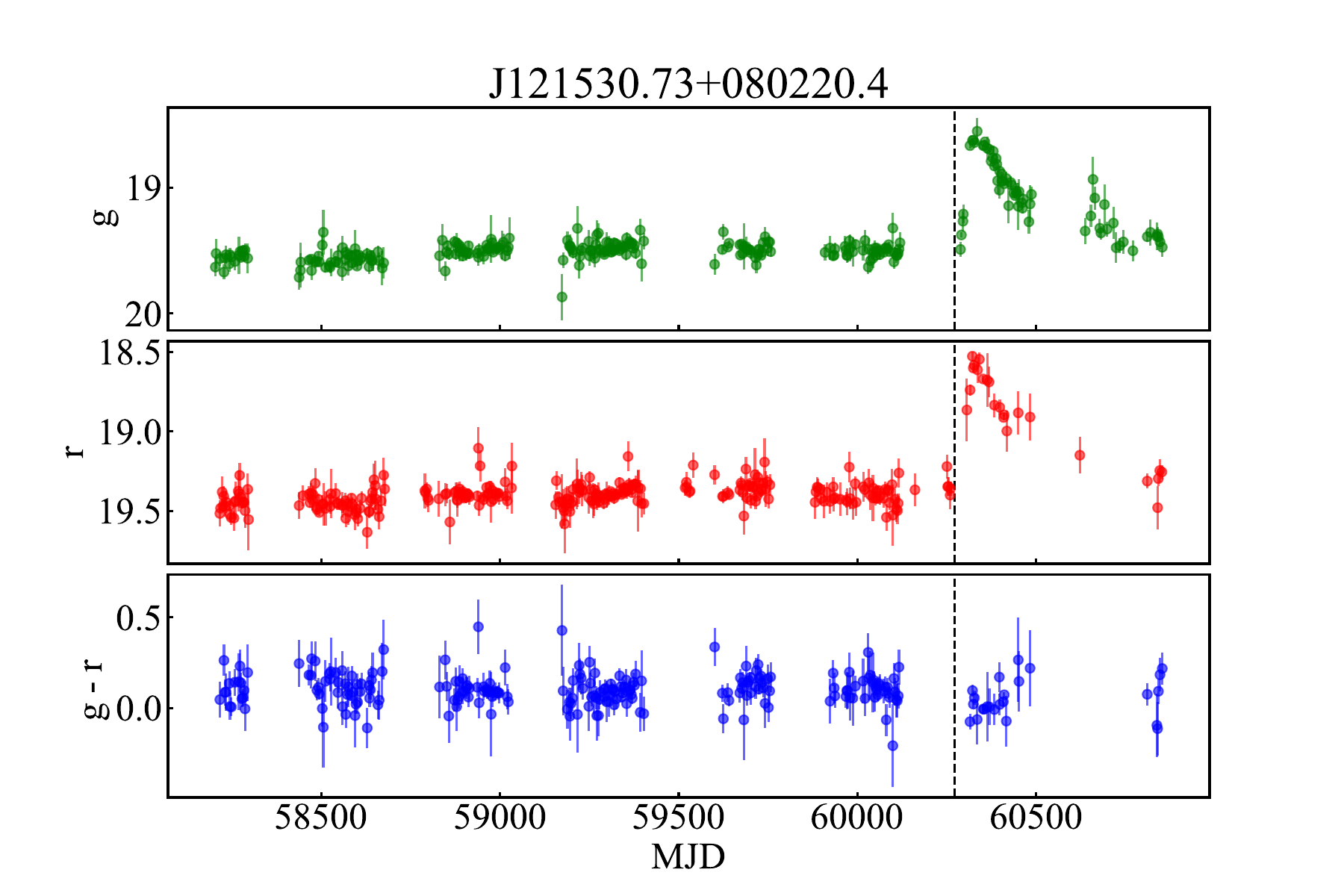}{0.5\textwidth}{}
    \fig{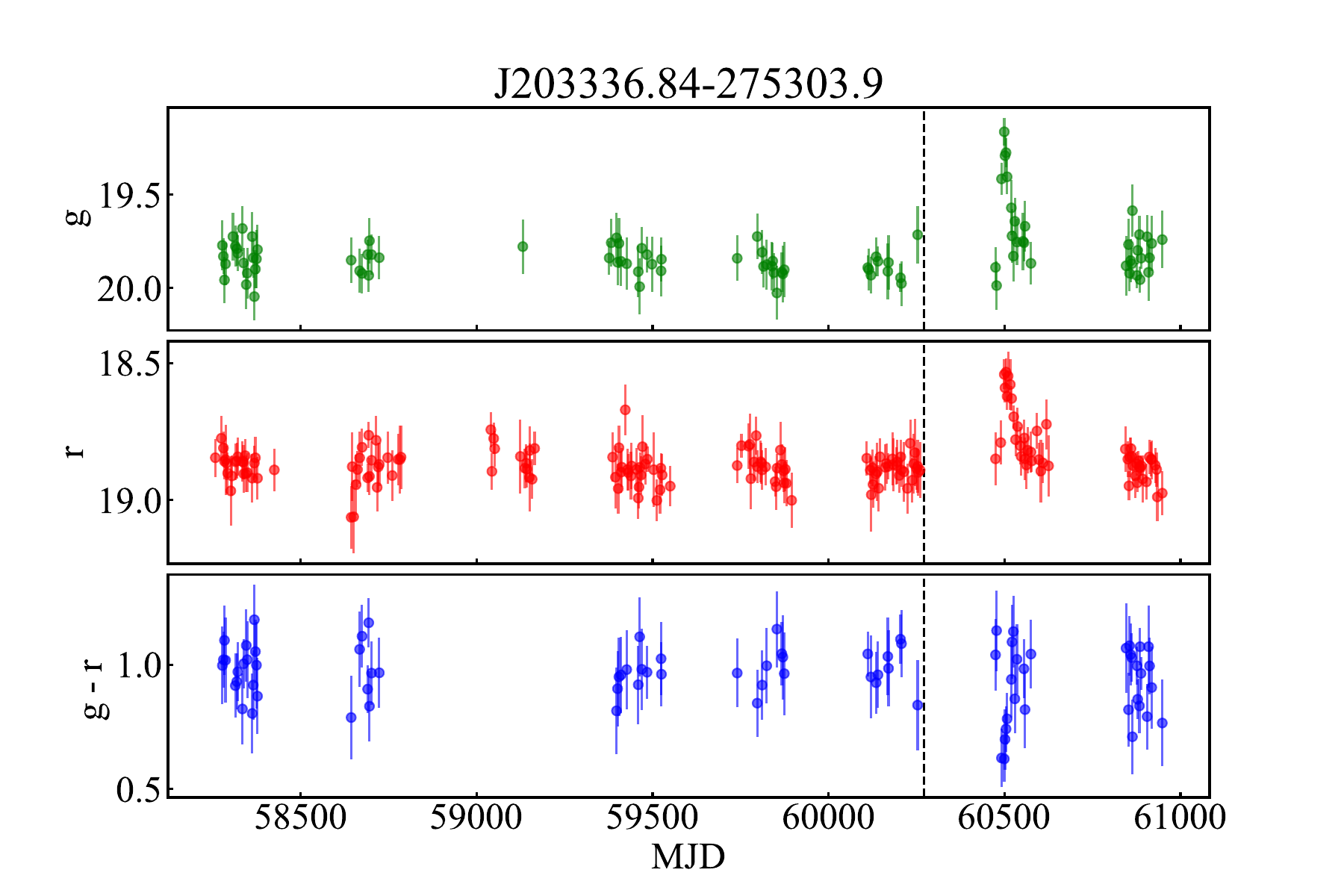}{0.5\textwidth}{}
  }
  \gridline{
    \fig{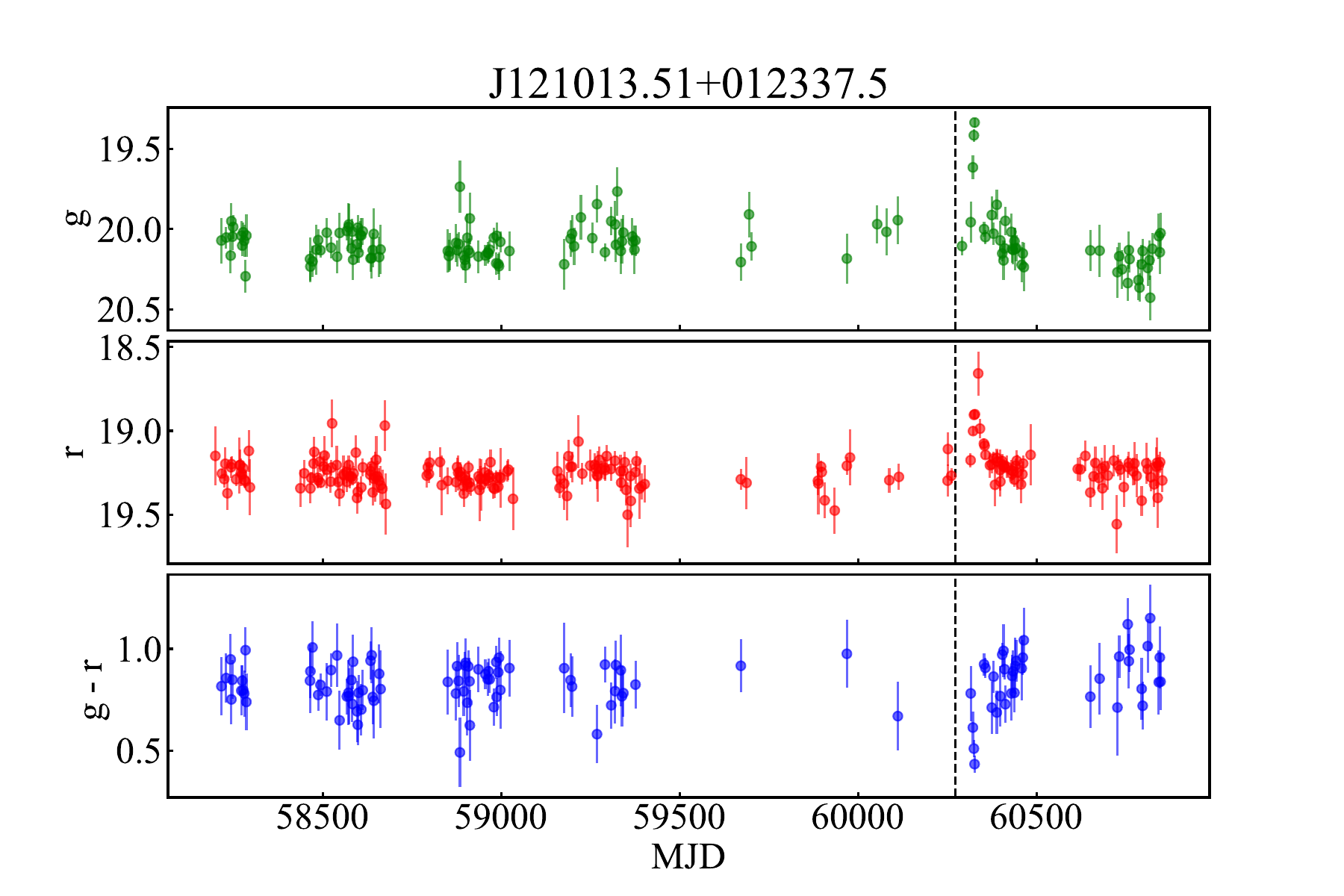}{0.5\textwidth}{}
    \fig{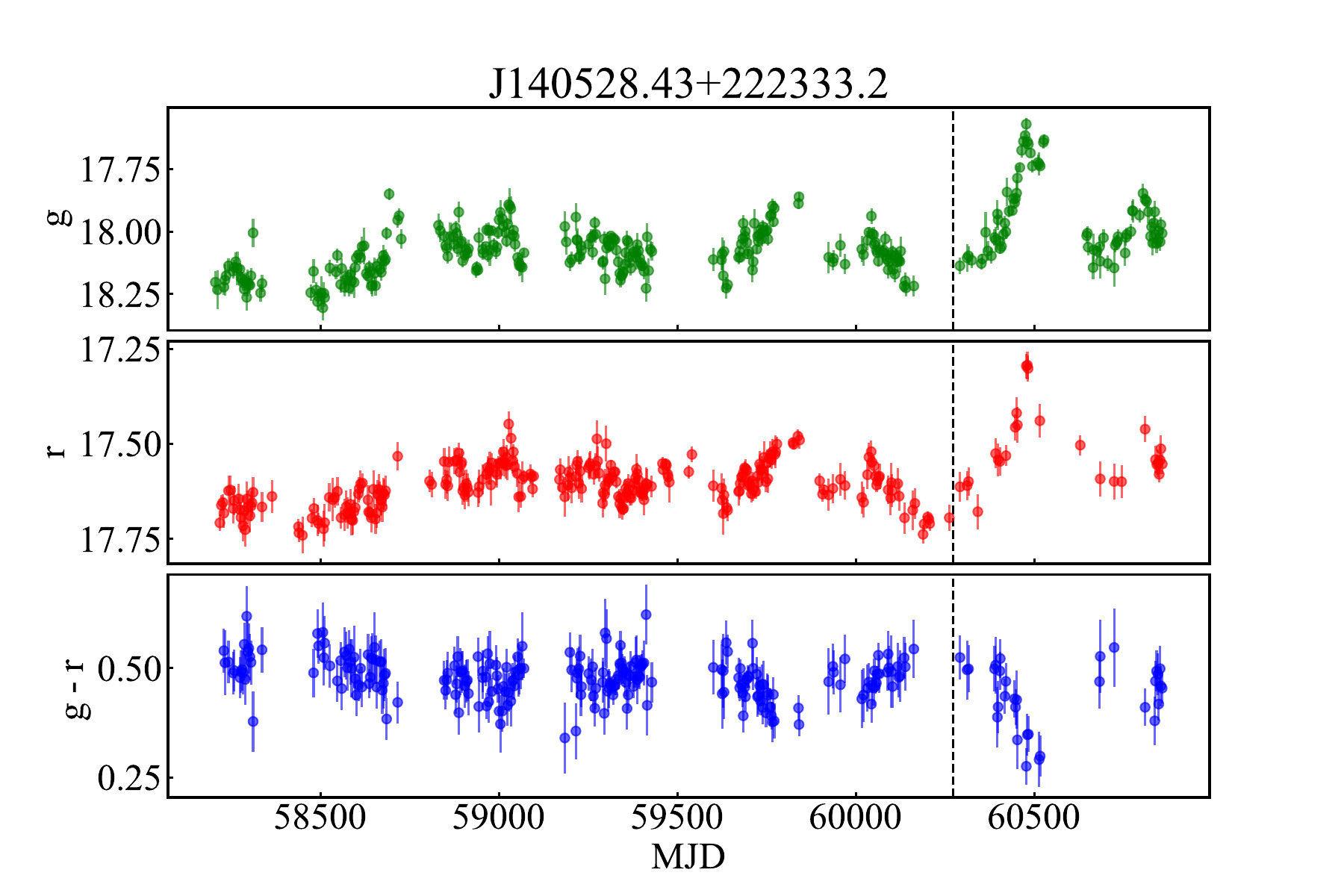}{0.5\textwidth}{}
  }
  \label{fig:Lightcurves}
  \caption{ZTF g-band photometry, r-band photometry, and g-r color for AGN flares that are selected as potential EM counterparts of GW231123.}
\end{figure*}

Applying the procedure described in Section~\ref{sec:method}, we find six AGN flares that could potentially be associated with GW231123. Based on the Gaussian process simulations, J140528.43+222333.2 reaches a significance level of 99.7\%, while the other five events exceed 99.9\%, relative to stochastic AGN variability. The key observational and derived properties of these potential counterparts are summarized in Table~\ref{tab:flares}, and their g- and r-band light curves are presented in Figure~\ref{fig:Lightcurves}.

\begin{deluxetable*}{ccccccccccc}[htb]
  \tablecaption{List of the 6 AGN flares that are selected as potential EM counterparts of GW231123.}
  \tablehead{
    \colhead{AGN name} & \colhead{RA} & \colhead{Dec} & \colhead{Redshift} & \colhead{$t_{\mathrm{exit}}$} & \colhead{$t_{\mathrm{g}}$} & \colhead{$t_{\mathrm{e}}$} & \colhead{$\log_{10} (E_{\mathrm{tot}})$} & \colhead{$\log_{10} (M_{\mathrm{SMBH}})$} & \colhead{$\rho$} & \colhead{$H$} \\
    & [deg] & [deg] & & [days] & [days] & [days] & [$\mathrm{erg}$] & [$M_{\odot}$] & [$ 10^{-10}\mathrm{g} \ \mathrm{cm}^{-3}$] & [$r_{\mathrm{g}}$]
  }
  \startdata
  J164911.49+492438.9 & 252.298 & 49.411 & 0.344 & 274.8 & 79.4 & 140.3 & 51.5 & 8.3 & 3.2 & 33\\
  J131824.85+253342.0 & 199.604 & 25.562 & 0.424* & 76.9 & 42.3 & 54.5 & 51.6 & (8.0) & 7.6 & (9.4)\\
  J121530.73+080220.4 & 183.878 & 8.039 & 0.802* & 28.9 & 8.4 & 74.0 & 52.4 & (8.0) & 241 & (3.5) \\
  J203336.84-275303.9 & 308.404 & -27.884 & 0.16 & 194.7 & 3.2 & 23.2 & 49.4 & 7.1 & 0.63 & 24\\
  J121013.51+012337.5 & 182.556 & 1.394 & 0.171 & 47.2 & 4.4 & 14.3 & 49.7 & 6.6 & 0.91 & 5.8\\
  J140528.43+222333.2 & 211.368 & 22.393 & 0.156 & 185.8 & 33.6 & 108.7 & 51.0 & 7.5 & 2.4 & 23
  \enddata
  \label{tab:flares}
  \tablecomments{Here $t_{\mathrm{exit}}$ denotes the time between the GW trigger and the flare peak, while $t_{\mathrm{g}}$ and $t_{\mathrm{e}}$ represent the Gaussian rise and exponential decay timescales, respectively. All timescales are measured in the rest frame of the AGN. Redshifts are spectroscopic for sources with available spectra, \added{taken from SDSS \citep{sdsscollaborationNineteenthDataRelease2025}, DESI \citep{desicollaborationDataRelease12025}}, or our SALT observations. Two AGNs lack spectroscopic observations, and we adopt redshifts from the Quaia catalog \citep{storey-fisherQuaiaGaiaunWISEQuasar2024} (asterisked values), which are derived from Gaia low-resolution BP/RP spectra, combined with unWISE infrared data, and calibrated using a machine-learning model trained on spectroscopic samples. A fiducial value of $\log_{10}(M_{\mathrm{SMBH}} / M_{\odot}) = 8$ is adopted (values are in parentheses) for these sources. The AGN disk density $\rho$ and the scale height are also listed. For sources with fiducial SMBH masses, the corresponding $H$ values are likewise given in parentheses.}

\end{deluxetable*}

Although no spectroscopic observations were taken during the flares, we conduct a search for archival spectra in the Sloan Digital Sky Survey (SDSS) DR19\footnote{\url{https://www.sdss.org/dr19/}} \citep{sdsscollaborationNineteenthDataRelease2025} and the Dark Energy Spectroscopic Instrument (DESI) DR1\footnote{\url{https://data.desi.lbl.gov/doc/releases/dr1/}} \citep{desicollaborationDataRelease12025}. This effort yields pre-flare spectra for three of the six candidates (see Appendix~ \ref{appendix:spectrum}). In addition, we obtain a new spectrum for J203336.84-275303.9 with the Robert Stobie Spectrograph \citep[RSS,][]{2003SPIE.4841.1463B} on the South African Large Telescope \citep[SALT,][]{2006SPIE.6267E..0ZB}, with an exposure time of 1200s. We fit all available spectra using PyQSOFit \citep{guoPyQSOFitPythonCode2018} to measure their broad-line properties. For most sources, we derive virial SMBH masses from the $H_\beta$ line profile based on the relation from \citet{hoREVISEDCALIBRATIONVIRIAL2015}. For J203336.84-275303.9, which does not show a clearly detectable $H_\beta$ line, we instead estimate the SMBH mass using the broad $H_\alpha$ emission line based on the relation from \citet{shenCOMPARINGSINGLEEPOCHVIRIAL2012}. These SMBH masses are also reported in Table~\ref{tab:flares}.

Table~\ref{tab:probability} lists the confidence levels, defined as the percentile contour of GW231123 at which each AGN is located. A lower value indicates a higher probability that the GW event occurred at the source location. Furthermore, in order to assess whether the flare and this GW event originate from the same source, we compute the odds ratio for every flare across all five waveforms in which the AGN lies within the 99\% confidence volume, following the procedure described in Section \ref{sec:odds}. \added{The confidence level and odds ratio should be regarded as complementary statistics. While the confidence level quantifies the localization consistency between the AGN and the GW event, the odds ratio additionally incorporates the number of AGNs located within the GW localization volume, the significance of the flare, and the expected mass distribution under different hypotheses.}

\begin{deluxetable*}{ccccccc}[htb]
  \tablecaption{Credible Interval (CI) and odds ratios for the six AGN flares spatially consistent with GW231123 under different waveform models }
  \tablehead{
    \colhead{AGN name} & & \colhead{IMRPhenomTPHM} & \colhead{IMRPhenomXO4a} & \colhead{IMRPhenomXPHM} & \colhead{NRSur7dq4} & \colhead{SEOBNRv5PHM}
  }
  \startdata
  J164911.49+492438.9 & 3D (2D) CI & 0.93 (0.98) & - (-) & - (-) & 0.98 (0.98) & 0.98 (-) \\
   & $\ln \mathcal{O}$ & 8.4 & - & - & 6.6 & 8.2 \\
   J131824.85+253342.0 & 3D (2D) CI& 0.64 (0.67) & 0.98 (0.68) & - (0.51) & 0.85 (0.09) & 0.72 (0.78) \\
    & $\ln \mathcal{O}$ & 10.3 & 6.8 & - & 9.2 & 9.9 \\
    J121530.73+080220.4 & 3D (2D) CI& - (0.85) & 0.98 (0.83) & - (0.91) & - (0.89) & - (0.91) \\
    &$\ln \mathcal{O}$& - & 8.8 & - & - & - \\
    J203336.84-275303.9 & 3D (2D) CI & - (-) & - (-) & 0.98 (0.98) & 0.98 (-) & - (-) \\
    & $\ln \mathcal{O}$ & - & - & 10.8 & 1.1 & - \\
    J121013.51+012337.5 & 3D (2D) CI & - (0.92) & - (0.93) & - (-) & 0.89 (0.97) & - (0.94) \\
    & $\ln \mathcal{O}$ & - & - & - & 9.1 & - \\
    J140528.43+222333.2 & 3D (2D) CI& - (-) & - (-) & - (-) & 0.91 (-) & - (-)\\
    & $\ln \mathcal{O}$ & - & - & - & 4.0 & - \\
  \enddata
  \label{tab:probability}
  \tablecomments{The confidence levels obtained under different waveform models are provided in the first row for each flare. Values outside parentheses denote the 3D results, whereas those in parentheses indicate the 2D results. Only cases with confidence levels below 0.99 are listed. For flares whose 3D confidence levels fall below this threshold, the odds ratios are computed following the procedure outlined in \citet{heTracingLightIdentification2025a}, and the logarithmic odds ratios ($\ln \mathcal{O}$) are shown in the second row for each flare.}
\end{deluxetable*}

An odds ratio greater than 150 ($\ln \mathcal{O} > 5$) is often interpreted as strong evidence for a physical association in a Bayesian framework \citep{kassBayesFactors1995}. In our analysis, five of the six candidates exceed this threshold for at least one waveform model (Table~\ref{tab:probability}), with the exception of J140528.43+222333.2. This is largely driven by the unusually large primary mass inferred for GW231123. In the association model, such high masses are not uncommon for BBHs embedded in AGN disks, whereas they are significantly less probable under the GW population distribution assumed in the coincidence model \citep[see figure~4 in][]{heTracingLightIdentification2025a}. As a result, events with large primary masses naturally lead to larger odds ratios in favor of the association hypothesis, and this effect becomes more pronounced for extreme events like GW231123.

The highest odds ratio is obtained for the AGN J203336.84-275303.9 under the IMRPhenomXPHM waveform model. This result is partly driven by the relatively small confidence volume associated with this waveform (see Figure~\ref{fig:GW-skymap}), which contains fewer AGNs and therefore increases the relative odds for any candidate flare located within it. In contrast, J140528.43+222333.2 has the lowest odds ratio, likely because its flare is less significant relative to its substantial intrinsic variability.

\begin{figure}[htb]
  \fig{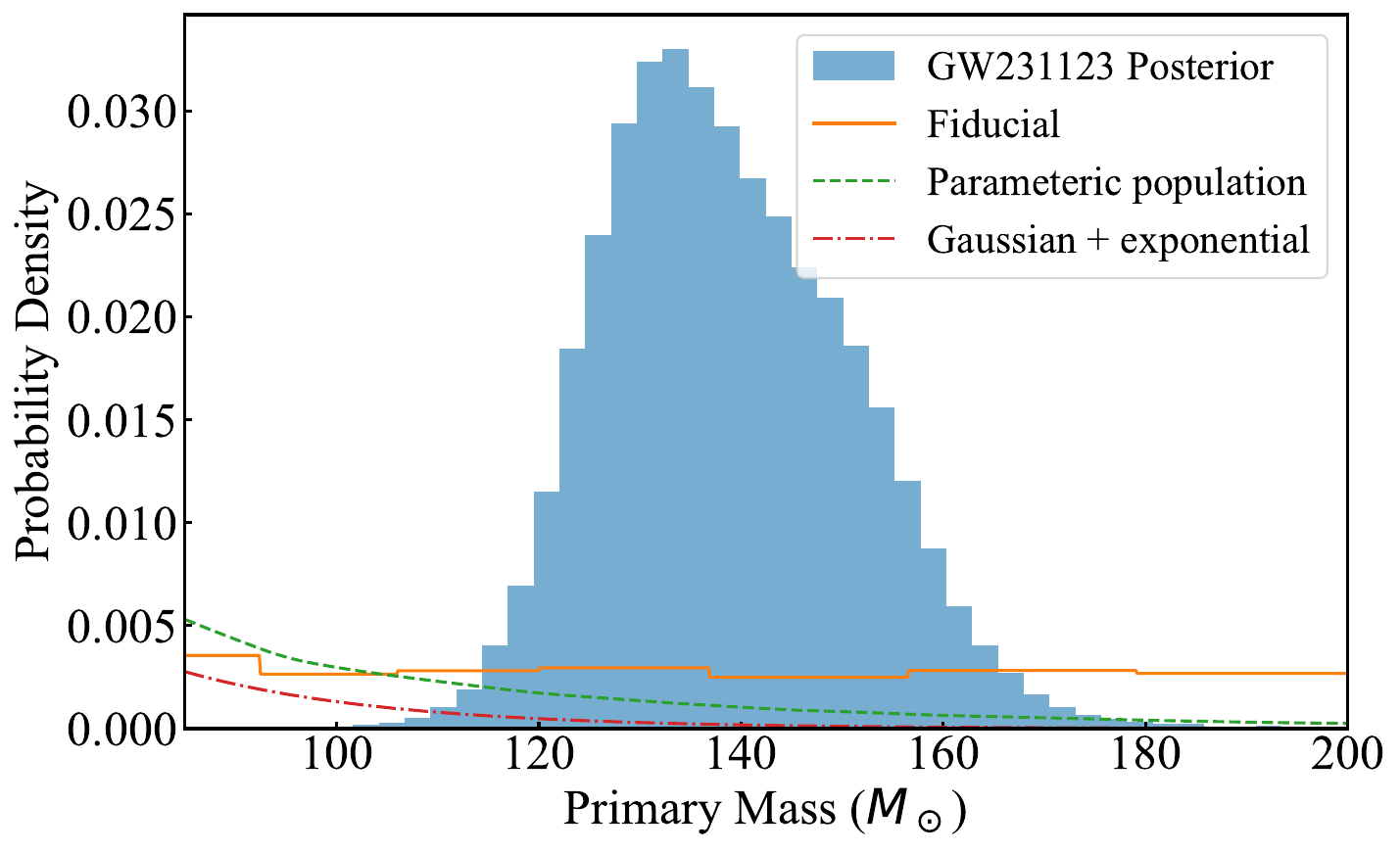}{0.47\textwidth}{}
  \caption{Comparison between the source-frame primary-mass posterior of GW231123 and the primary mass priors adopted in the odds ratio calculation.}
  \label{fig:mass-priors}
\end{figure}

\added{To assess the sensitivity of the odds ratios to the assumed BBH mass distribution, we repeated the calculation using two additional AGN-motivated source-frame primary mass priors\citep{liComparingHierarchicalBlack2023,afrozPhaseSpaceBinary2025}. These priors assign different probabilities in the high mass regime relevant for GW231123, as illustrated in Figure~\ref{fig:mass-priors}. For each candidate, we adopt the waveform model that gives the largest odds ratio in Table~\ref{tab:probability}. The results are summarized in Table~\ref{tab:probability_prior}. Although the absolute values of the odds ratios change with the adopted mass prior, all candidates except J140528.43+222333.2 remain above the threshold of $\ln \mathcal{O}>5$ for all three choices. The relative ranking is also broadly unchanged, with J203336.84-275303.9 and J131824.85+253342.0 consistently among the highest-ranked candidates. This indicates that the identification of the most promising candidates is not driven by a particular choice of BBH mass prior.} Since at most one flare can be physically associated with a given GW event, the odds ratios presented here should primarily be interpreted as a statistical metric for ranking candidates rather than as direct association probabilities. Alternative approaches to quantify such association have also been proposed \citep[e.g., ][]{cabreraMultimessengerConstraintsLIGO2026}.

\begin{deluxetable*}{ccccc}[htb]
  \tablecaption{Odds ratios ($\ln\mathcal{O}$) obtained using different source-frame primary mass priors.}
  \tablehead{
    \colhead{AGN name} & \colhead{waveform} & \colhead{Fiducial} & \colhead{Parametric population} & \colhead{Gaussian + exponential }
  }
  \startdata
  J164911.49+492438.9 & IMRPhenomTPHM & 8.4 & 7.2 & 5.5 \\
  J131824.85+253342.0 & IMRPhenomTPHM & 10.3 & 9.1 & 7.6\\
  J121530.73+080220.4 & IMRPhenomXO4a & 8.8 & 7.5 & 5.9  \\
  J203336.84-275303.9 & IMRPhenomXPHM & 10.8 & 9.2 & 7.2 \\
  J121013.51+012337.5 & NRSur7dq4 & 9.1 & 8.0 & 6.4 \\
  J140528.43+222333.2 & NRSur7dq4 & 4.0 & 3.0 & 1.5 \\
  \enddata
  \label{tab:probability_prior}
  \tablecomments{For each candidate, we use the waveform model that gives the largest odds ratio in Table~\ref{tab:probability}. The fiducial prior follows \citet{vaccaroImpactGasHardening2024}, the parametric population prior follows \citet{liComparingHierarchicalBlack2023}, and the Gaussian + exponential prior follows \citet{afrozPhaseSpaceBinary2025}.}
\end{deluxetable*}

To test the impact of the assumed cosmology, we recomputed the confidence levels and odds ratios assuming the SH0ES value of $H_0$ \citep{riessComprehensiveMeasurementLocal2022}. The results are summarized in Table~\ref{tab:probability_SH0ES}. A few candidate pairs originally located near the boundary of the localization region fall outside the 99\% credible volume under this assumption. For the remaining candidates, the values change only slightly and the relative ranking remains largely unaffected.

\begin{deluxetable*}{ccccccc}[htb]
  \tablecaption{Same as Table \ref{tab:probability}, but assuming the SH0ES cosmology }
  \label{tab:probability_SH0ES}
  \tablehead{
    \colhead{AGN name} & & \colhead{IMRPhenomTPHM} & \colhead{IMRPhenomXO4a} & \colhead{IMRPhenomXPHM} & \colhead{NRSur7dq4} & \colhead{SEOBNRv5PHM}
  }
  \startdata
  J164911.49+492438.9 & 3D (2D) CI & 0.91 (0.98) & - (-) & - (-) & 0.97 (0.98) & 0.98 (-) \\
   & $\ln \mathcal{O}$ & 8.0 & - & - & 6.4 & 8.0 \\
   J131824.85+253342.0 & 3D (2D) CI& 0.69 (0.68) & - (0.69) & - (0.52) & 0.77 (0.09) & 0.70 (0.78) \\
   & $\ln \mathcal{O}$ & 9.3 & - & - & 9.5 & 10.1 \\
   J121530.73+080220.4 & 3D (2D) CI& - (0.87) & 0.92 (0.83) & - (0.91) & - (0.89) & - (0.91) \\
    &$\ln \mathcal{O}$& - & 9.9 & - & - & - \\
    J203336.84-275303.9 & 3D (2D) CI & - (-) & - (-) & - (0.98) & 0.98 (-) & - (-) \\
    & $\ln \mathcal{O}$ & - & - & - & 0.3 & - \\
    J121013.51+012337.5 & 3D (2D) CI & - (0.92) & - (0.93) & - (-) & 0.88 (0.97) & - (0.94) \\
    & $\ln \mathcal{O}$ & - & - & - & 8.5 & - \\
    J140528.43+222333.2 & 3D (2D) CI& - (-) & - (-) & - (-) & 0.90 (-) & - (-)\\
    & $\ln \mathcal{O}$ & - & - & - & 3.0 & - \\
  \enddata
\end{deluxetable*}

\section{Discussion\label{sec:discussion}}
To explore a potential physical origin for our candidate flares, we adopt a model of a BBH merger occurring within an AGN disk, based on the framework discussed by \citet{grahamCandidateElectromagneticCounterpart2020,grahamLightDarkSearching2023}. In this scenario, the post-merger remnant recoils with a kick velocity ($v_{\mathrm{k}}$), estimated for GW231123 to be $v_{\mathrm{k}} = 967^{+829}_{-857} \mathrm{km \ s^{-1}} $ \citep{theligoscientificcollaborationGW231123BinaryBlack2025}. Gas initially bound to the binary is dragged along with the remnant, colliding with the surrounding disk gas. This interaction generates an initial, shock-powered flare on a characteristic timescale $t_{\mathrm{bound}} = GM_{\mathrm{BBH}}/v_{\mathrm{k}}^3$ \citep{mckernanRampressureStrippingKicked2019}, which can be parameterized as
\begin{equation}
  \label{eq:1}
  t_{\mathrm{bound}} \sim 20 \ \mathrm{days} \left( \frac{M_{\mathrm{BBH}}}{100M_{\odot}} \right) \left( \frac{v_{\mathrm{k}}}{200 \mathrm{km \ s^{-1}}} \right)^{-3}.
\end{equation}

This prompt emission mechanism, while likely a lower-luminosity precursor to the main flare, provides a means of interpreting the observed delay times. For GW231123, using its remnant mass of $\sim 200M_{\odot}$ and the median kick velocity of $\sim 1000 \mathrm{km \ s^{-1}}$, the expected delay is extremely short, $t_{\mathrm{bound}} \sim 0.4 \mathrm{days}$. This provides a plausible explanation for the nearly immediate onset of flares such as J131824.85+253342.0, J121530.73+080220.4, and J121013.51+012337.5. However, the uncertainty in the kick velocity is substantial and we cannot rule out the flares that occur after a longer delay.

Once the merger remnant receives a recoil kick, it may detach from the gas that was originally gravitationally bound to it. As the remnant moves through the surrounding AGN disk, the nearby gas is swept up and accelerated, forming a shocked Bondi tail \citep{edgarReviewBondiHoyle2004,antoniEvolutionBinariesGaseous2019}. This tail both exerts a dynamical drag on the black hole and feeds gas back onto it through accretion. The resulting Bondi-Hoyle-Lyttleton (BHL) luminosity could be expressed as $L_{\mathrm{BHL}}=\eta \dot{M}_{\mathrm{BHL}}c^{2}$, where $\eta \sim 0.1$ is the radiative efficiency and the mass accretion rate is given by
\begin{equation}
  \label{eq:2}
  \dot{M}_{\mathrm{BHL}} = \frac{4\pi G^{2}M_{\mathrm{BBH}}^{2}\rho}{v_{\mathrm{rel}}^{3}}.
\end{equation}
Here, $\rho$ denotes the local gas density of the AGN disk and $v_{\mathrm{rel}}$ is the relative velocity between the remnant and the ambient gas. We assume that the sound speed $c_{\mathrm{s}}$ in the AGN disk gas is less than $v_{\mathrm{k}}$, so $v_{\mathrm{rel}} \sim v_{\mathrm{k}}$.

In this simple scenario, the emission is expected to exhibit an approximately constant color \citep{grahamCandidateElectromagneticCounterpart2020}. To test this prediction, we quantify the color evolution for each flare by defining the pre-GW g-r distribution as a baseline and computing the significance of post-GW deviations. We find that J121530.73+080220.4 exhibits statistically consistent constant colors after the GW event, while the other sources show significant decreases in g-r at the $> 3 \sigma$ level relative to their historical distributions. The most prominent case is J164911.49+492438.9, which exhibits a decrease in g-r exceeding $5\sigma$. The color variations seen in these candidates may reflect more complex physical conditions beyond the scope of this simple treatment.

To further constrain the physical condition in the disk, we use the observed flare luminosities to infer the corresponding gas densities, assuming that the measured flare luminosity directly reflects the BHL accretion luminosity described above. The total energy released by each flare ($E_{\mathrm{tot}}$) is estimated by assuming a roughly constant spectral flux density across the g- and B-band frequency domain and applying the typical quasar bolometric correction factor for the B band from \citet{durasUniversalBolometricCorrections2020}.

Following the approach of \citet{grahamLightDarkSearching2023}, we assume that the recoil kick is directed perpendicular to the AGN disk midplane, corresponding to the fastest possible escape path for the merger remnant. Considering a vertically stratified disk with a Gaussian density profile, the time required for the remnant to exit the optically thick region of the disk is given by

\begin{equation}
  \label{eq:3} t_{\mathrm{exit}}=\frac{H\sqrt{2\ln(\tau_{\mathrm{mp}})}}{v_{\mathrm{k}}},
\end{equation}
where $\tau_{\mathrm{mp}}$ is the optical depth at the merger point. The peak of the resulting flare is expected to occur on a timescale $\approx t_{\mathrm{exit}}$, which is defined as the time it takes for the remnant to reach the $\tau=1$ surface from the merger location. Adopting a typical value of $\tau_{\mathrm{mp}} \sim 10^{4.5}$ \citep{cabreraSearchingElectromagneticEmission2024a}, we can estimate the disk scale height $H$ for the AGN disk. As shown in Table \ref{tab:flares}, the combination of the derived values $\rho$ and $H$ is consistent with the predictions of the AGN disk structure models \citep{Sirko2003,Thompson2005}.

Given the deep gravitational potential well of the SMBH, the recoil velocity of the merged remnant is expected to be insufficient for it to escape the AGN disk. Consequently, a recurrent flare may occur when the remnant reencounters the disk on a timescale of $1.6\mathrm{yr}(M_{\mathrm{SMBH}}/10^{8}M_{\odot})(r/10^{3}r_{\mathrm{g}})^{3/2}$ \citep{grahamCandidateElectromagneticCounterpart2020}, where $r$ denotes the distance between the SMBH and the BBH merger site. Therefore, continued monitoring for these AGNs is essential to test the BBH-origin hypothesis for these flares. Detection of a second flare would provide a valuable opportunity to constrain the merger location within the AGN disk through the measured recurrence interval.

Moreover, the flare produced by the kicked BH in an AGN disk is off-center, which would lead to asymmetric illumination of the broad-line region and consequently asymmetric broad-line profiles in the spectrum \citep{mckernanRampressureStrippingKicked2019}. Although timely spectroscopy is not available for the initial flare, future spectroscopic observations of any recurrent flares could provide decisive evidence to confirm their physical origin.

Other types of transient, such as tidal disruption events \citep[TDE, ][]{karasEnhancedActivityMassive2007,ryuInplaneTidalDisruption2024}, supernovae \citep[SNe, ][]{zhuThermonuclearExplosionsAccretioninduced2021,liCorecollapseSupernovaExplosions2023}, or kilonovae \citep{Ren2022}, could also produce similar flares in AGN disks and thus represent possible false positives. Since we have already excluded known transients in Section~\ref{sec:method}, no classifications for these flares are reported in the Transient Name Server\footnote{\url{https://www.wis-tns.org}}. According to the Automatic Learning for the Rapid Classification of Events (ALeRCE) broker light curve classifier \citep{sanchez-saezAlertClassificationALeRCE2021} of the ZTF alert stream\footnote{\url{https://alerce.online}}, J203336.84-275303.9 is classified as a supernova, J121013.51+012337.5 remains unclassified, and the remaining flares are classified as AGN. This classifier relies solely on photometric data and does not include a class corresponding to BBH mergers, but its results offer valuable insight.

To further investigate the nature of these flare candidates, several follow-up efforts will be essential:
\begin{enumerate}
  \item Continuous optical monitoring: Long-term, high-cadence observations with time-domain facilities like the Wide Field Survey Telescope \citep{wangScience25meterWide2023} and the Vera C. Rubin observatory \citep{thelsstsciencecollaborationsScienceDrivenOptimizationLSST2017,ivezicLSSTScienceDrivers2019,andreoniRubinToO20242024} will be crucial for detecting the predicted repeat flares and distinguishing genuine counterparts from stochastic AGN variability.
  \item Rapid multi-wavelength and spectroscopic follow-up: Coordinated multi-band observations, together with timely spectroscopic observation during the flare phase, are vital for confirming a possible BBH merger origin and distinguishing it from other types of transients. In particular, identifying asymmetric broad-line features, while ruling out typical spectral signatures of SNe or TDEs, would provide strong diagnostic evidence \citep{cabreraSearchingElectromagneticEmission2024a}.
  \item High-energy observations: X-ray and $\gamma$-ray monitoring with facilities such as the Fermi Gamma-ray Space Telescope \citep{thompsonFermiGammaRaySpace2023} or the Einstein Probe \citep{yuanEinsteinProbeMission2022}  can provide additional constraints on the physical environment. Even non-detections would offer valuable information for discriminating between different emission models.
\end{enumerate}

\section{Conclusions\label{sec:conclusion}}
We have conducted a systematic search for potential EM counterparts to GW231123, the most massive BBH merger detected to date. By crossmatching the GW localization with the AGN Flare Catalog from ZTF DR23, we identify six AGN flares that are spatially and temporally consistent with this extraordinary event. Their distinctive variability relative to the host AGN baseline makes them promising candidates for further investigation.

To assess the likelihood of a physical association, we calculate the odds ratios between the association model and the coincidence model for each candidate. Most candidates show strong evidence for association under this Bayesian framework, which is primarily driven by the large primary mass of GW231123. Furthermore, by applying a simple Bondi accretion model, we show that we could constrain the physical environment of the host AGN disk. However, the absence of contemporaneous spectroscopic data means that the physical origin of these flares remains inconclusive.

Among the six plausible candidates, at most one could represent the true EM counterpart of GW231123, while the others are likely unrelated background events. Future observation will be crucial to identify the genuine counterpart. A confirmed association would not only validate the AGN-assisted merger scenario but also mark an important step forward in the era of multi-messenger astronomy.

\begin{acknowledgements}

  This work is supported by the National Natural Science Foundation of China (grant Nos. 12325301, 12273035 and 12405075), Strategic Priority Research Program of the Chinese Academy of Science (grant No. XDB0550300), the National Key R\&D Program of China (grant Nos. 2021YFC2203102, 2022YFC2204602, and 2024YFC2207500), the Science Research Grants from the China Manned Space Project (grant No. CMS-CSST-2021-B01), the 111 Project for ``Observational and Theoretical Research on Dark Matter and Dark Energy" (grant No. B23042), and Cyrus Chun Ying Tang Foundations. 

  This paper makes use of data from the Southern African Large Telescope (SALT) via Rutgers University program 2025-1-MLT-001 (PI: S.W.J.). The ZTF forced-photometry service was funded under the Heising-Simons Foundation grant \#12540303 (PI: Graham).

  This research used data obtained with the Sloan Digital Sky Survey (SDSS) and Dark Energy Spectroscopic Instrument (DESI). SDSS is managed by the Astrophysical Research Consortium for the participating institutions of the SDSS Collaboration, including Caltech, The Carnegie Institution for Science, Chilean National Time Allocation Committee (CNTAC) ratified researchers, The Flatiron Institute, the Gotham Participation Group, Harvard University, Heidelberg University, The Johns Hopkins University, L'Ecole polytechnique fédérale de Lausanne (EPFL), Leibniz-Institut für Astrophysik Potsdam (AIP), Max-Planck-Institut für Astronomie (MPIA Heidelberg), Max-Planck-Institut für Extraterrestrische Physik (MPE), Nanjing University, National Astronomical Observatories of China (NAOC), New Mexico State University, The Ohio State University, Pennsylvania State University, Smithsonian Astrophysical Observatory, Space Telescope Science Institute (STScI), the Stellar Astrophysics Participation Group, Universidad Nacional Autónoma de México, University of Arizona, University of Colorado Boulder, University of Illinois at Urbana-Champaign, University of Toronto, University of Utah, University of Virginia, Yale University, and Yunnan University. 
  
  DESI construction and operations is managed by the Lawrence Berkeley National Laboratory. This material is based upon work supported by the U.S. Department of Energy, Office of Science, Office of High-Energy Physics, under Contract No. DE–AC02–05CH11231, and by the National Energy Research Scientific Computing Center, a DOE Office of Science User Facility under the same contract. Additional support for DESI was provided by the U.S. National Science Foundation (NSF), Division of Astronomical Sciences under Contract No. AST-0950945 to the NSF’s National Optical-Infrared Astronomy Research Laboratory; the Science and Technology Facilities Council of the United Kingdom; the Gordon and Betty Moore Foundation; the Heising-Simons Foundation; the French Alternative Energies and Atomic Energy Commission (CEA); the National Council of Humanities, Science and Technology of Mexico (CONAHCYT); the Ministry of Science and Innovation of Spain (MICINN), and by the DESI Member Institutions: www.desi.lbl.gov/collaborating-institutions. The DESI collaboration is honored to be permitted to conduct scientific research on I’oligam Du’ag (Kitt Peak), a mountain with particular significance to the Tohono O’odham Nation. Any opinions, findings, and conclusions or recommendations expressed in this material are those of the author(s) and do not necessarily reflect the views of the U.S. National Science Foundation, the U.S. Department of Energy, or any of the listed funding agencies.

\end{acknowledgements}

\software{
  numpy \citep{harrisArrayProgrammingNumPy2020}, matplotlib \citep{hunterMatplotlib2DGraphics2007}, ligo.skymap \citep{singerRapidBayesianPosition2016}, astropy \citep{astropycollaborationAstropyCommunityPython2013,astropycollaborationAstropyProjectBuilding2018,astropycollaborationAstropyProjectSustaining2022}, pandas \citep{mckinneyDataStructuresStatistical2010}, PyQSOFit \citep{guoPyQSOFitPythonCode2018}
}

\appendix
\section{Spectra of Candidate Counterparts of GW231123\label{appendix:spectrum}}

\begin{figure*}[htb]
  \centering
  \gridline{
    \fig{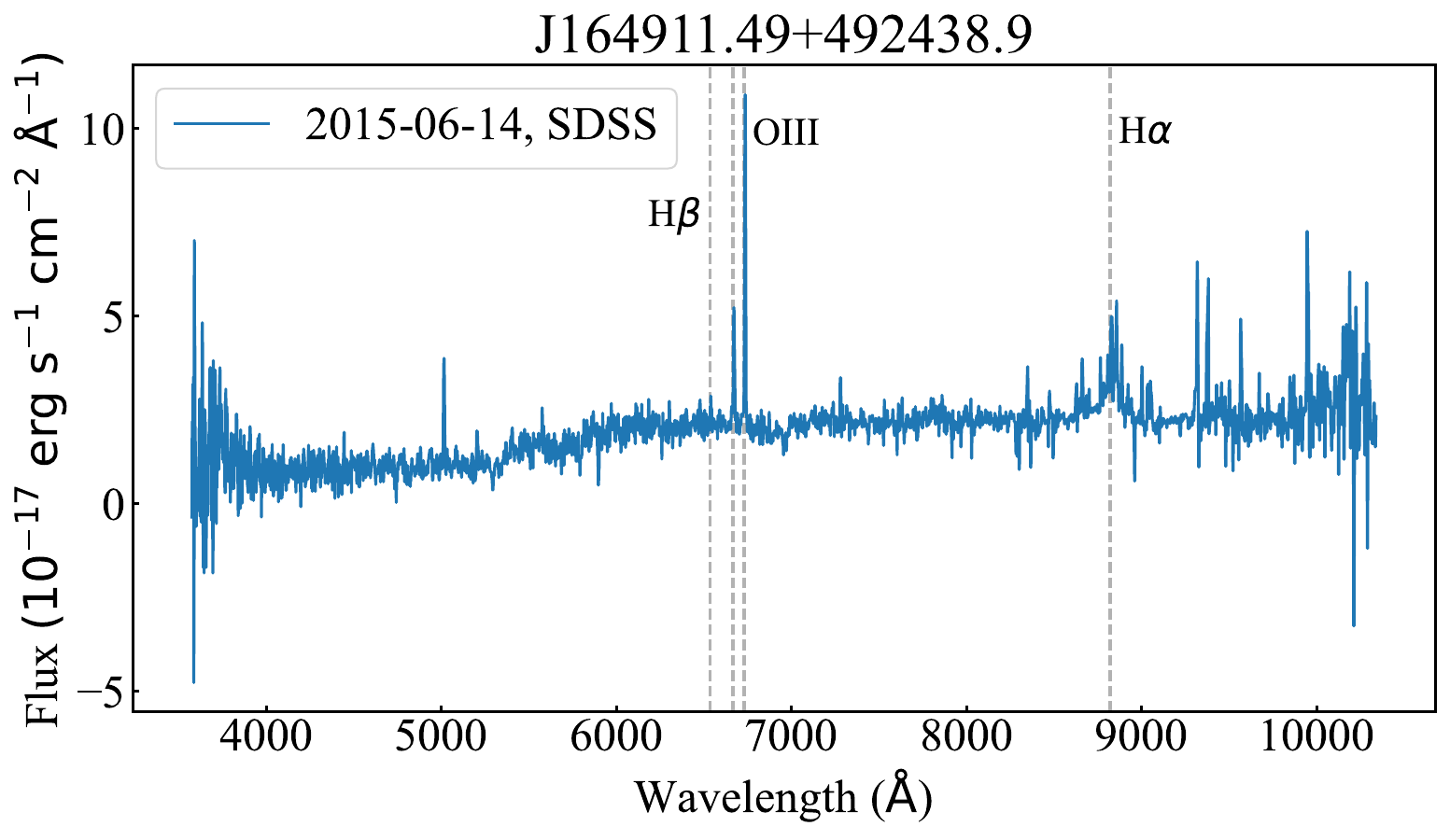}{0.5\textwidth}{}
    \fig{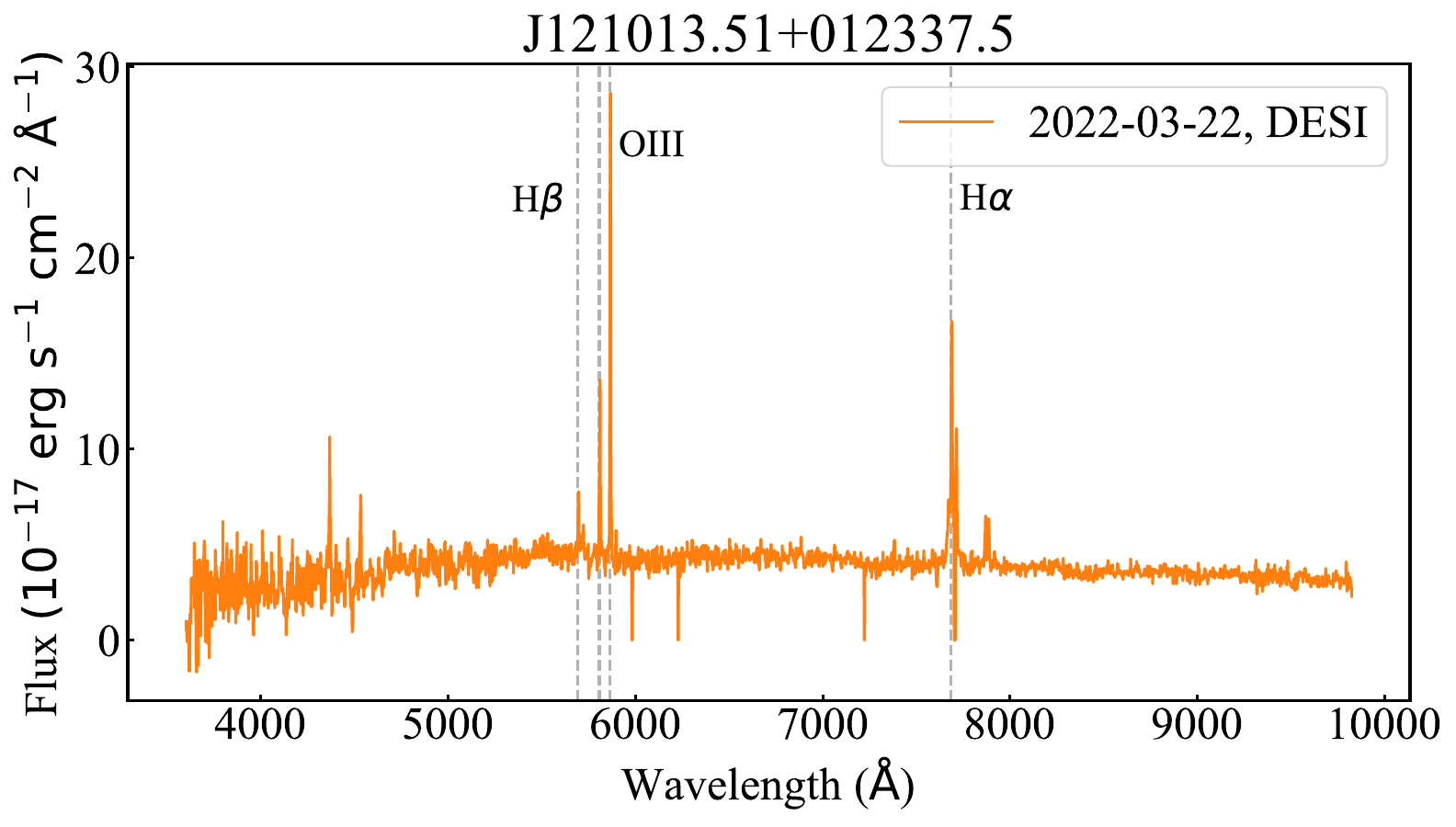}{0.5\textwidth}{}
  }
  \gridline{
    \fig{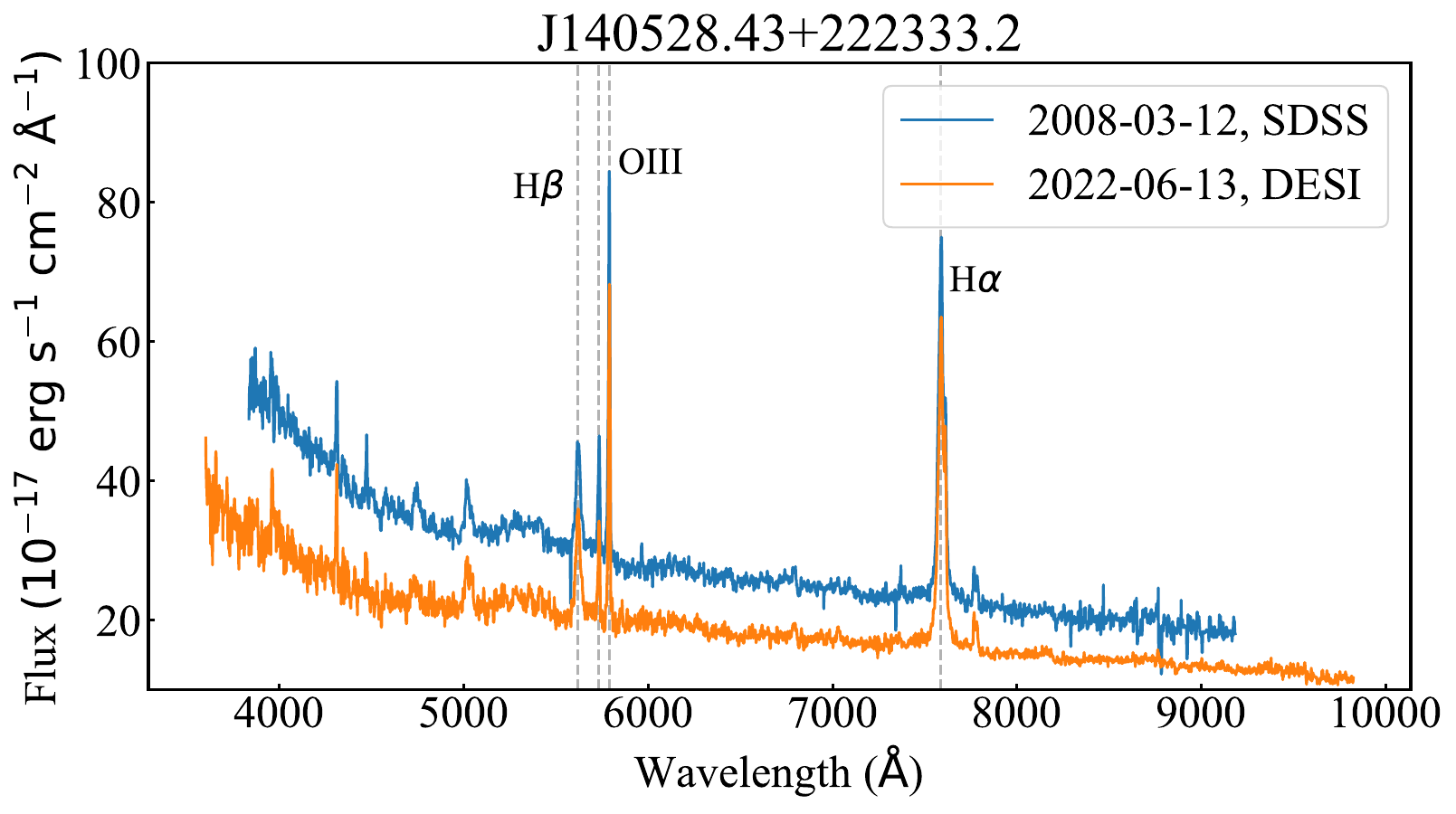}{0.5\textwidth}{}
    \fig{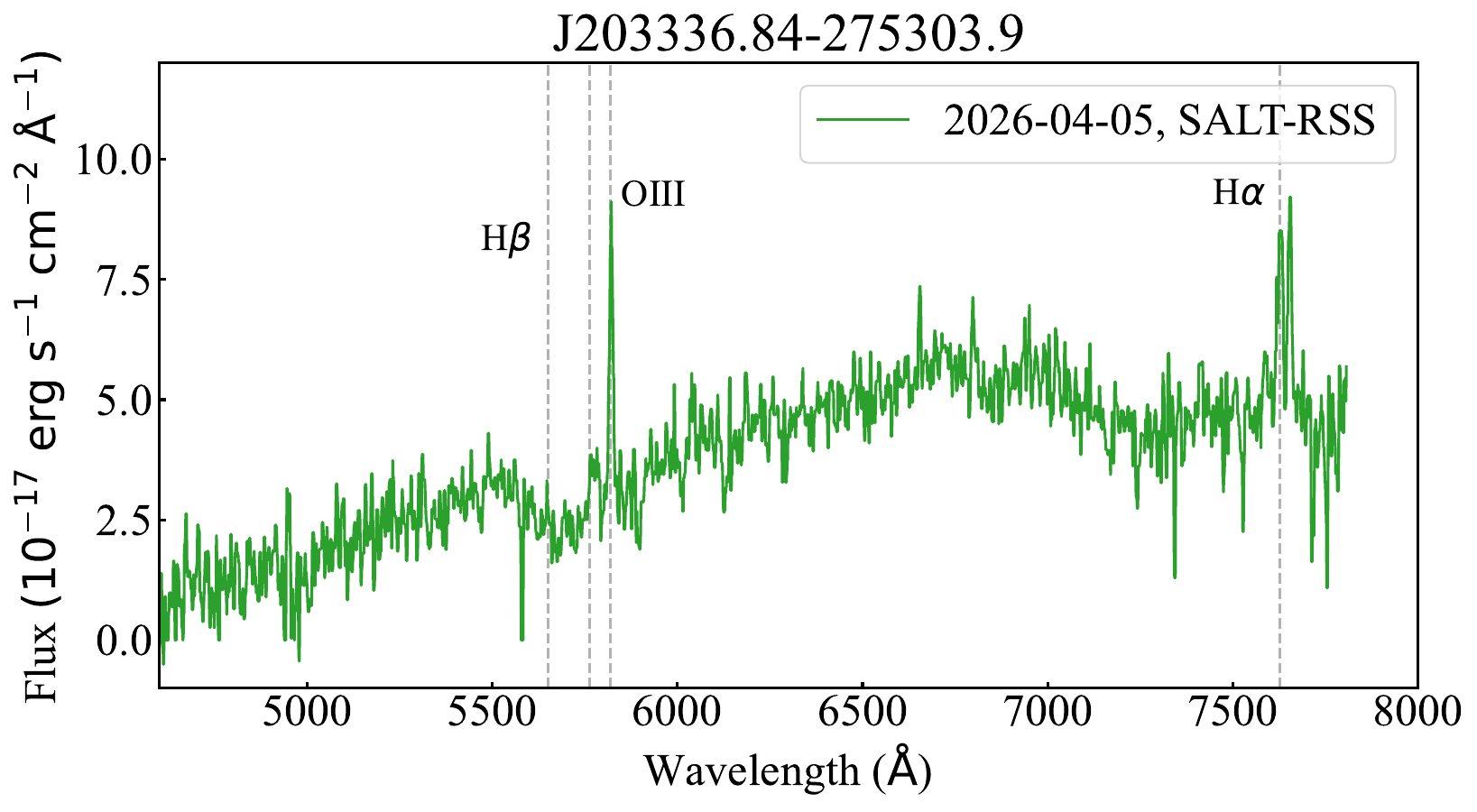}{0.5\textwidth}{}
  }
  \caption{The spectra of the AGNs potentially associated with GW231123. \added{The expected locations of the H$\alpha$, H$\beta$, and O III lines are marked by dashed gray lines. All spectra are shown in the observed frame.} \label{fig:spectra}}  
\end{figure*}

Here, we present archival and newly obtained spectra for the candidate counterparts of GW231123. The sources J164911.49+492438.9, J121013.51+012337.5 and J140528.43+222333.2 have archival spectra. Specifically, J164911.49+492438.9 has one spectrum from SDSS, J121013.51+012337.5 has one from DESI, and J140528.43+222333.2 has spectra from both SDSS and DESI. All archival spectra were obtained prior to the onset of the corresponding flares. In addition, we include a newly obtained spectrum of J203336.84-275303.9, acquired with RSS on the SALT after our initial analysis. All these spectra are shown in Figure~\ref{fig:spectra}.

\section{Environmental Effects in the AGN Disk\label{appendix:environment}}

For a BBH embedded in an AGN disk, the observed GW signal can be affected not only by the cosmological redshift of the host AGN, but also by environmental effects associated with the motion of the BBH around the central SMBH and the gravitational potential of the SMBH. These effects modify the mapping from intrinsic source-frame parameters to the effective detector-frame quantities.

Following \citet{mortonGW190521BinaryBlack2023}, we assume that the BBH is on a circular orbit around a non-spinning SMBH, and the relativistic redshift and gravitational redshift are given by
\begin{equation}
  z_{\mathrm{rel}} = \gamma (1+v\cos(\theta)) - 1, 
  \quad z_{\mathrm{grav}} = \sqrt{1-\frac{R_{\rm S}}{r}} - 1.
\end{equation}
Here $\gamma = (1- v^2)^{-1/2}$ is the Lorentz factor, $v=[2(r/R_{\rm S}) - 1]^{-1/2}$ is the magnitude of the velocity, $\theta$ is the viewing angle between the velocity and the line of sight in the observer frame, $R_{\rm S}$ is the Schwarzschild radius of the SMBH, and $r$ is the distance between the BBH and the SMBH. 

The effective primary mass and luminosity distance inferred from the GW signal are then related to the intrinsic source-frame quantities by
\begin{equation}
  M_1^{\mathrm{eff}} = M_1 (1+z)(1+z_{\mathrm{rel}})(1+z_{\mathrm{grav}}), \quad
  D_L^{\mathrm{eff}} = D_L(z)(1+z_{\mathrm{rel}})^2(1+z_{\mathrm{grav}}).
\end{equation}
Here $M_1$ is the source-frame primary mass, $z$ is the cosmological redshift of the host AGN, and $D_L(z)$ is the corresponding cosmological luminosity distance.

In practice, we sample the environmental variables $r$ and $\theta$ in addition to other parameters. We adopt a radial prior $p(r)\propto r$ over $1<r/R_{\rm S}<3000$, and a uniform prior in $\cos\theta$. For each sample, the above equations are used to map the intrinsic source-frame quantities to the corresponding effective detector-frame quantities.

\bibliography{main}

@article{aghanimPlanck2018Results2020,
  title = {Planck 2018 Results - {{VI}}. {{Cosmological}} Parameters},
  author = {Aghanim, N. and Akrami, Y. and Ashdown, M. and Aumont, J. and Baccigalupi, C. and Ballardini, M. and Banday, A. J. and Barreiro, R. B. and Bartolo, N. and Basak, S. and Battye, R. and Benabed, K. and Bernard, J.-P. and Bersanelli, M. and Bielewicz, P. and Bock, J. J. and Bond, J. R. and Borrill, J. and Bouchet, F. R. and Boulanger, F. and Bucher, M. and Burigana, C. and Butler, R. C. and Calabrese, E. and Cardoso, J.-F. and Carron, J. and Challinor, A. and Chiang, H. C. and Chluba, J. and Colombo, L. P. L. and Combet, C. and Contreras, D. and Crill, B. P. and Cuttaia, F. and de Bernardis, P. and de Zotti, G. and Delabrouille, J. and Delouis, J.-M. and Valentino, E. Di and Diego, J. M. and Dor{\'e}, O. and Douspis, M. and Ducout, A. and Dupac, X. and Dusini, S. and Efstathiou, G. and Elsner, F. and En{\ss}lin, T. A. and Eriksen, H. K. and Fantaye, Y. and Farhang, M. and Fergusson, J. and {Fernandez-Cobos}, R. and Finelli, F. and Forastieri, F. and Frailis, M. and Fraisse, A. A. and Franceschi, E. and Frolov, A. and Galeotta, S. and Galli, S. and Ganga, K. and {G{\'e}nova-Santos}, R. T. and Gerbino, M. and Ghosh, T. and {Gonz{\'a}lez-Nuevo}, J. and G{\'o}rski, K. M. and Gratton, S. and Gruppuso, A. and Gudmundsson, J. E. and Hamann, J. and Handley, W. and Hansen, F. K. and Herranz, D. and Hildebrandt, S. R. and Hivon, E. and Huang, Z. and Jaffe, A. H. and Jones, W. C. and Karakci, A. and Keih{\"a}nen, E. and Keskitalo, R. and Kiiveri, K. and Kim, J. and Kisner, T. S. and Knox, L. and Krachmalnicoff, N. and Kunz, M. and {Kurki-Suonio}, H. and Lagache, G. and Lamarre, J.-M. and Lasenby, A. and Lattanzi, M. and Lawrence, C. R. and Jeune, M. Le and Lemos, P. and Lesgourgues, J. and Levrier, F. and Lewis, A. and Liguori, M. and Lilje, P. B. and Lilley, M. and Lindholm, V. and {L{\'o}pez-Caniego}, M. and Lubin, P. M. and Ma, Y.-Z. and {Mac{\'i}as-P{\'e}rez}, J. F. and Maggio, G. and Maino, D. and Mandolesi, N. and Mangilli, A. and {Marcos-Caballero}, A. and Maris, M. and Martin, P. G. and Martinelli, M. and {Mart{\'i}nez-Gonz{\'a}lez}, E. and Matarrese, S. and Mauri, N. and McEwen, J. D. and Meinhold, P. R. and Melchiorri, A. and Mennella, A. and Migliaccio, M. and Millea, M. and Mitra, S. and {Miville-Desch{\^e}nes}, M.-A. and Molinari, D. and Montier, L. and Morgante, G. and Moss, A. and Natoli, P. and {N{\o}rgaard-Nielsen}, H. U. and Pagano, L. and Paoletti, D. and Partridge, B. and Patanchon, G. and Peiris, H. V. and Perrotta, F. and Pettorino, V. and Piacentini, F. and Polastri, L. and Polenta, G. and Puget, J.-L. and Rachen, J. P. and Reinecke, M. and Remazeilles, M. and Renzi, A. and Rocha, G. and Rosset, C. and Roudier, G. and {Rubi{\~n}o-Mart{\'i}n}, J. A. and {Ruiz-Granados}, B. and Salvati, L. and Sandri, M. and Savelainen, M. and Scott, D. and Shellard, E. P. S. and Sirignano, C. and Sirri, G. and Spencer, L. D. and Sunyaev, R. and {Suur-Uski}, A.-S. and Tauber, J. A. and Tavagnacco, D. and Tenti, M. and Toffolatti, L. and Tomasi, M. and Trombetti, T. and Valenziano, L. and Valiviita, J. and Tent, B. Van and Vibert, L. and Vielva, P. and Villa, F. and Vittorio, N. and Wandelt, B. D. and Wehus, I. K. and White, M. and White, S. D. M. and Zacchei, A. and Zonca, A.},
  year = {2020},
  month = sep,
  journal = {Astronomy \& Astrophysics},
  volume = {641},
  pages = {A6},
  publisher = {EDP Sciences},
  issn = {0004-6361, 1432-0746},
  doi = {10.1051/0004-6361/201833910},
  urldate = {2025-01-16},
  copyright = {{\copyright} ESO 2020},
  langid = {english}
}

@article{antoniEvolutionBinariesGaseous2019,
  title = {The {{Evolution}} of {{Binaries}} in a {{Gaseous Medium}}: {{Three-dimensional Simulations}} of {{Binary Bondi}}--{{Hoyle}}--{{Lyttleton Accretion}}},
  shorttitle = {The {{Evolution}} of {{Binaries}} in a {{Gaseous Medium}}},
  author = {Antoni, Andrea and MacLeod, Morgan and {Ramirez-Ruiz}, Enrico},
  year = {2019},
  month = oct,
  journal = {The Astrophysical Journal},
  volume = {884},
  number = {1},
  pages = {22},
  publisher = {The American Astronomical Society},
  issn = {0004-637X},
  doi = {10.3847/1538-4357/ab3466},
  urldate = {2024-07-09},
  langid = {english}
}

@article{ashtonCurrentObservationsAre2021,
  title = {Current Observations Are Insufficient to Confidently Associate the Binary Black Hole Merger {{GW190521}} with {{AGN J124942}}.3+344929},
  author = {Ashton, Gregory and Ackley, Kendall and Hernandez, Ignacio Maga{\~n}a and Piotrzkowski, Brandon},
  year = {2021},
  month = dec,
  journal = {Classical and Quantum Gravity},
  volume = {38},
  number = {23},
  eprint = {2009.12346},
  primaryclass = {astro-ph},
  pages = {235004},
  issn = {0264-9381, 1361-6382},
  doi = {10.1088/1361-6382/ac33bb},
  urldate = {2024-05-11},
  archiveprefix = {arXiv},
  langid = {american}
}

@article{astropycollaborationAstropyCommunityPython2013,
  title = {Astropy: {{A}} Community {{Python}} Package for Astronomy},
  shorttitle = {Astropy},
  author = {{Astropy Collaboration} and Robitaille, Thomas P. and Tollerud, Erik J. and Greenfield, Perry and Droettboom, Michael and Bray, Erik and Aldcroft, Tom and Davis, Matt and Ginsburg, Adam and {Price-Whelan}, Adrian M. and Kerzendorf, Wolfgang E. and Conley, Alexander and Crighton, Neil and Barbary, Kyle and Muna, Demitri and Ferguson, Henry and Grollier, Fr{\'e}d{\'e}ric and Parikh, Madhura M. and Nair, Prasanth H. and Unther, Hans M. and Deil, Christoph and Woillez, Julien and Conseil, Simon and Kramer, Roban and Turner, James E. H. and Singer, Leo and Fox, Ryan and Weaver, Benjamin A. and Zabalza, Victor and Edwards, Zachary I. and Azalee Bostroem, K. and Burke, D. J. and Casey, Andrew R. and Crawford, Steven M. and Dencheva, Nadia and Ely, Justin and Jenness, Tim and Labrie, Kathleen and Lim, Pey Lian and Pierfederici, Francesco and Pontzen, Andrew and Ptak, Andy and Refsdal, Brian and Servillat, Mathieu and Streicher, Ole},
  year = {2013},
  month = oct,
  journal = {Astronomy and Astrophysics},
  volume = {558},
  pages = {A33},
  issn = {0004-6361},
  doi = {10.1051/0004-6361/201322068},
  urldate = {2025-04-18}
}

@article{astropycollaborationAstropyProjectBuilding2018,
  title = {The {{Astropy Project}}: {{Building}} an {{Open-science Project}} and {{Status}} of the v2.0 {{Core Package}}},
  shorttitle = {The {{Astropy Project}}},
  author = {{Astropy Collaboration} and {Price-Whelan}, A. M. and Sip{\H o}cz, B. M. and G{\"u}nther, H. M. and Lim, P. L. and Crawford, S. M. and Conseil, S. and Shupe, D. L. and Craig, M. W. and Dencheva, N. and Ginsburg, A. and VanderPlas, J. T. and Bradley, L. D. and {P{\'e}rez-Su{\'a}rez}, D. and {de Val-Borro}, M. and Aldcroft, T. L. and Cruz, K. L. and Robitaille, T. P. and Tollerud, E. J. and Ardelean, C. and Babej, T. and Bach, Y. P. and Bachetti, M. and Bakanov, A. V. and Bamford, S. P. and Barentsen, G. and Barmby, P. and Baumbach, A. and Berry, K. L. and Biscani, F. and Boquien, M. and Bostroem, K. A. and Bouma, L. G. and Brammer, G. B. and Bray, E. M. and Breytenbach, H. and Buddelmeijer, H. and Burke, D. J. and Calderone, G. and Cano Rodr{\'i}guez, J. L. and Cara, M. and Cardoso, J. V. M. and Cheedella, S. and Copin, Y. and Corrales, L. and Crichton, D. and D'Avella, D. and Deil, C. and Depagne, {\'E}. and Dietrich, J. P. and Donath, A. and Droettboom, M. and Earl, N. and Erben, T. and Fabbro, S. and Ferreira, L. A. and Finethy, T. and Fox, R. T. and Garrison, L. H. and Gibbons, S. L. J. and Goldstein, D. A. and Gommers, R. and Greco, J. P. and Greenfield, P. and Groener, A. M. and Grollier, F. and Hagen, A. and Hirst, P. and Homeier, D. and Horton, A. J. and Hosseinzadeh, G. and Hu, L. and Hunkeler, J. S. and Ivezi{\'c}, {\v Z}. and Jain, A. and Jenness, T. and Kanarek, G. and Kendrew, S. and Kern, N. S. and Kerzendorf, W. E. and Khvalko, A. and King, J. and Kirkby, D. and Kulkarni, A. M. and Kumar, A. and Lee, A. and Lenz, D. and Littlefair, S. P. and Ma, Z. and Macleod, D. M. and Mastropietro, M. and McCully, C. and Montagnac, S. and Morris, B. M. and Mueller, M. and Mumford, S. J. and Muna, D. and Murphy, N. A. and Nelson, S. and Nguyen, G. H. and Ninan, J. P. and N{\"o}the, M. and Ogaz, S. and Oh, S. and Parejko, J. K. and Parley, N. and Pascual, S. and Patil, R. and Patil, A. A. and Plunkett, A. L. and Prochaska, J. X. and Rastogi, T. and Reddy Janga, V. and Sabater, J. and Sakurikar, P. and Seifert, M. and Sherbert, L. E. and {Sherwood-Taylor}, H. and Shih, A. Y. and Sick, J. and Silbiger, M. T. and Singanamalla, S. and Singer, L. P. and Sladen, P. H. and Sooley, K. A. and Sornarajah, S. and Streicher, O. and Teuben, P. and Thomas, S. W. and Tremblay, G. R. and Turner, J. E. H. and Terr{\'o}n, V. and {van Kerkwijk}, M. H. and {de la Vega}, A. and Watkins, L. L. and Weaver, B. A. and Whitmore, J. B. and Woillez, J. and Zabalza, V. and {Astropy Contributors}},
  year = {2018},
  month = sep,
  journal = {The Astronomical Journal},
  volume = {156},
  pages = {123},
  publisher = {IOP},
  issn = {0004-6256},
  doi = {10.3847/1538-3881/aabc4f},
  urldate = {2025-04-18}
}

@article{astropycollaborationAstropyProjectSustaining2022,
  title = {The {{Astropy Project}}: {{Sustaining}} and {{Growing}} a {{Community-oriented Open-source Project}} and the {{Latest Major Release}} (v5.0) of the {{Core Package}}},
  shorttitle = {The {{Astropy Project}}},
  author = {{Astropy Collaboration} and {Price-Whelan}, Adrian M. and Lim, Pey Lian and Earl, Nicholas and Starkman, Nathaniel and Bradley, Larry and Shupe, David L. and Patil, Aarya A. and Corrales, Lia and Brasseur, C. E. and N{\"o}the, Maximilian and Donath, Axel and Tollerud, Erik and Morris, Brett M. and Ginsburg, Adam and Vaher, Eero and Weaver, Benjamin A. and Tocknell, James and Jamieson, William and {van Kerkwijk}, Marten H. and Robitaille, Thomas P. and Merry, Bruce and Bachetti, Matteo and G{\"u}nther, H. Moritz and Aldcroft, Thomas L. and {Alvarado-Montes}, Jaime A. and Archibald, Anne M. and B{\'o}di, Attila and Bapat, Shreyas and Barentsen, Geert and Baz{\'a}n, Juanjo and Biswas, Manish and Boquien, M{\'e}d{\'e}ric and Burke, D. J. and Cara, Daria and Cara, Mihai and Conroy, Kyle E. and Conseil, Simon and Craig, Matthew W. and Cross, Robert M. and Cruz, Kelle L. and D'Eugenio, Francesco and Dencheva, Nadia and Devillepoix, Hadrien A. R. and Dietrich, J{\"o}rg P. and Eigenbrot, Arthur Davis and Erben, Thomas and Ferreira, Leonardo and {Foreman-Mackey}, Daniel and Fox, Ryan and Freij, Nabil and Garg, Suyog and Geda, Robel and Glattly, Lauren and Gondhalekar, Yash and Gordon, Karl D. and Grant, David and Greenfield, Perry and Groener, Austen M. and Guest, Steve and Gurovich, Sebastian and Handberg, Rasmus and Hart, Akeem and {Hatfield-Dodds}, Zac and Homeier, Derek and Hosseinzadeh, Griffin and Jenness, Tim and Jones, Craig K. and Joseph, Prajwel and Kalmbach, J. Bryce and Karamehmetoglu, Emir and Ka{\l}uszy{\'n}ski, Miko{\l}aj and Kelley, Michael S. P. and Kern, Nicholas and Kerzendorf, Wolfgang E. and Koch, Eric W. and Kulumani, Shankar and Lee, Antony and Ly, Chun and Ma, Zhiyuan and MacBride, Conor and Maljaars, Jakob M. and Muna, Demitri and Murphy, N. A. and Norman, Henrik and O'Steen, Richard and Oman, Kyle A. and Pacifici, Camilla and Pascual, Sergio and {Pascual-Granado}, J. and Patil, Rohit R. and Perren, Gabriel I. and Pickering, Timothy E. and Rastogi, Tanuj and Roulston, Benjamin R. and Ryan, Daniel F. and Rykoff, Eli S. and Sabater, Jose and Sakurikar, Parikshit and Salgado, Jes{\'u}s and Sanghi, Aniket and Saunders, Nicholas and Savchenko, Volodymyr and Schwardt, Ludwig and {Seifert-Eckert}, Michael and Shih, Albert Y. and Jain, Anany Shrey and Shukla, Gyanendra and Sick, Jonathan and Simpson, Chris and Singanamalla, Sudheesh and Singer, Leo P. and Singhal, Jaladh and Sinha, Manodeep and Sip{\H o}cz, Brigitta M. and Spitler, Lee R. and Stansby, David and Streicher, Ole and {\v S}umak, Jani and Swinbank, John D. and Taranu, Dan S. and Tewary, Nikita and Tremblay, Grant R. and {de Val-Borro}, Miguel and Van Kooten, Samuel J. and Vasovi{\'c}, Zlatan and Verma, Shresth and {de Miranda Cardoso}, Jos{\'e} Vin{\'i}cius and Williams, Peter K. G. and Wilson, Tom J. and Winkel, Benjamin and {Wood-Vasey}, W. M. and Xue, Rui and Yoachim, Peter and Zhang, Chen and Zonca, Andrea and {Astropy Project Contributors}},
  year = {2022},
  month = aug,
  journal = {The Astrophysical Journal},
  volume = {935},
  pages = {167},
  publisher = {IOP},
  issn = {0004-637X},
  doi = {10.3847/1538-4357/ac7c74},
  urldate = {2025-04-18}
}

@misc{bartosAccretionAllYou2025,
  title = {Accretion Is {{All You Need}}: {{Black Hole Spin Alignment}} in {{Merger GW231123 Indicates Accretion Pathway}}},
  shorttitle = {Accretion Is {{All You Need}}},
  author = {Bartos, Imre and Haiman, Zoltan},
  year = {2025},
  month = aug,
  number = {arXiv:2508.08558},
  eprint = {2508.08558},
  primaryclass = {astro-ph},
  publisher = {arXiv},
  doi = {10.48550/arXiv.2508.08558},
  urldate = {2025-08-13},
  archiveprefix = {arXiv},
  langid = {american}
}

@article{bartosRapidBrightStellarmass2017,
  title = {Rapid and {{Bright Stellar-mass Binary Black Hole Mergers}} in {{Active Galactic Nuclei}}},
  author = {Bartos, Imre and Kocsis, Bence and Haiman, Zolt{\'a}n and M{\'a}rka, Szabolcs},
  year = {2017},
  month = jan,
  journal = {The Astrophysical Journal},
  volume = {835},
  number = {2},
  pages = {165},
  publisher = {The American Astronomical Society},
  issn = {0004-637X},
  doi = {10.3847/1538-4357/835/2/165},
  urldate = {2025-01-17},
  langid = {english}
}

@article{beckPS1STRMNeuralNetwork2021,
  title = {{{PS1-STRM}}: Neural Network Source Classification and Photometric Redshift Catalogue for {{PS1}} 3{$\pi$} {{DR1}}},
  shorttitle = {{{PS1-STRM}}},
  author = {Beck, R{\'o}bert and Szapudi, Istv{\'a}n and Flewelling, Heather and Holmberg, Conrad and Magnier, Eugene and Chambers, Kenneth C.},
  year = {2021},
  month = jan,
  journal = {Monthly Notices of the Royal Astronomical Society},
  volume = {500},
  pages = {1633--1644},
  publisher = {OUP},
  issn = {0035-8711},
  doi = {10.1093/mnras/staa2587},
  urldate = {2025-05-11},
  langid = {american}
}

@article{cabreraSearchingElectromagneticEmission2024a,
  title = {Searching for Electromagnetic Emission in an {{AGN}} from the Gravitational Wave Binary Black Hole Merger Candidate {{S230922g}}},
  author = {Cabrera, Tom{\'a}s and Palmese, Antonella and Hu, Lei and O'Connor, Brendan and Ford, K. E. Saavik and McKernan, Barry and Andreoni, Igor and Ahumada, Tom{\'a}s and Amsellem, Ariel and Busmann, Malte and Clark, Peter and Coughlin, Michael W. and Dadiani, Ekaterine and Diaz, Veronica and Graham, Matthew J. and Gruen, Daniel and Kunnumkai, Keerthi and Postiglione, Jake and Riffeser, Arno and Sommer, Julian S. and Valdes, Francisco},
  year = {2024},
  month = dec,
  journal = {Physical Review D},
  volume = {110},
  number = {12},
  pages = {123029},
  publisher = {American Physical Society},
  doi = {10.1103/PhysRevD.110.123029},
  urldate = {2025-10-07}
}

@article{chenElectromagneticCounterpartsPowered2024,
  title = {Electromagnetic {{Counterparts Powered}} by {{Kicked Remnants}} of {{Black Hole Binary Mergers}} in {{AGN Disks}}},
  author = {Chen, Ken and Dai, Zi-Gao},
  year = {2024},
  month = jan,
  journal = {The Astrophysical Journal},
  volume = {961},
  number = {2},
  pages = {206},
  publisher = {The American Astronomical Society},
  issn = {0004-637X},
  doi = {10.3847/1538-4357/ad0dfd},
  urldate = {2025-01-17},
  langid = {english}
}

@article{chenStandardSirenCosmological2022,
  title = {A Standard Siren Cosmological Measurement from the Potential {{GW190521}} Electromagnetic Counterpart {{ZTF19abanrhr}}},
  author = {Chen, Hsin-Yu and Haster, Carl-Johan and Vitale, Salvatore and Farr, Will M. and Isi, Maximiliano},
  year = {2022},
  month = jun,
  journal = {Monthly Notices of the Royal Astronomical Society},
  volume = {513},
  pages = {2152--2157},
  publisher = {OUP},
  issn = {0035-8711},
  doi = {10.1093/mnras/stac989},
  urldate = {2025-01-16}
}

@misc{croonCanStellarPhysics2025,
  title = {Can Stellar Physics Explain {{GW231123}}?},
  author = {Croon, Djuna and Sakstein, Jeremy and Gerosa, Davide},
  year = {2025},
  month = aug,
  number = {arXiv:2508.10088},
  eprint = {2508.10088},
  primaryclass = {astro-ph},
  publisher = {arXiv},
  doi = {10.48550/arXiv.2508.10088},
  urldate = {2025-08-18},
  archiveprefix = {arXiv},
  langid = {american}
}

@misc{delfaveroProspectsFormationGW2311232025,
  title = {Prospects for the Formation of {{GW231123}} from the {{AGN}} Channel},
  author = {Delfavero, V. and Ray, S. and Cook, H. E. and Nathaniel, K. and McKernan, B. and Ford, K. E. S. and Postiglione, J. and McPike, E. and O'Shaughnessy, R.},
  year = {2025},
  month = aug,
  number = {arXiv:2508.13412},
  eprint = {2508.13412},
  primaryclass = {gr-qc},
  publisher = {arXiv},
  doi = {10.48550/arXiv.2508.13412},
  urldate = {2025-08-20},
  archiveprefix = {arXiv},
  langid = {american}
}

@misc{desicollaborationDataRelease12025,
  title = {Data {{Release}} 1 of the {{Dark Energy Spectroscopic Instrument}}},
  author = {{DESI Collaboration} and {Abdul-Karim}, M. and Adame, A. G. and Aguado, D. and Aguilar, J. and Ahlen, S. and Alam, S. and Aldering, G. and Alexander, D. M. and Alfarsy, R. and Allen, L. and Prieto, C. Allende and Alves, O. and Anand, A. and Andrade, U. and Armengaud, E. and Avila, S. and Aviles, A. and Awan, H. and Bailey, S. and Lizancos, A. Baleato and Ballester, O. and Bault, A. and Bautista, J. and BenZvi, S. and e Silva, L. Beraldo and {Bermejo-Climent}, J. R. and Beutler, F. and Bianchi, D. and Blake, C. and Blum, R. and Bolton, A. S. and Bonici, M. and Brieden, S. and Brodzeller, A. and Brooks, D. and {Buckley-Geer}, E. and Burtin, E. and Canning, R. and Rosell, A. Carnero and Carr, A. and Carrilho, P. and Casas, L. and Castander, F. J. and Cereskaite, R. and {Cervantes-Cota}, J. L. and Chaussidon, E. and {Chaves-Montero}, J. and Chen, S. and Chen, X. and Claybaugh, T. and Cole, S. and Cooper, A. P. and Cousinou, M.-C. and Cuceu, A. and Davis, T. M. and Dawson, K. S. and de Belsunce, R. and de la Cruz, R. and de la Macorra, A. and de Mattia, A. and Deiosso, N. and Costa, J. Della and Demina, R. and Demirbozan, U. and DeRose, J. and Dey, A. and Dey, B. and Ding, J. and Ding, Z. and Doel, P. and Douglass, K. and Dowicz, M. and Ebina, H. and Edelstein, J. and Eisenstein, D. J. and Elbers, W. and Emas, N. and Escoffier, S. and Fagrelius, P. and Fan, X. and Fanning, K. and Fawcett, V. A. and {Fern{\'a}ndez-Garc{\'i}a}, E. and Ferraro, S. and Findlay, N. and {Font-Ribera}, A. and {Forero-Romero}, J. E. and {Forero-S{\'a}nchez}, D. and Frenk, C. S. and G{\"a}nsicke, B. T. and Galbany, L. and {Garc{\'i}a-Bellido}, J. and {Garcia-Quintero}, C. and Garrison, L. H. and Gazta{\~n}aga, E. and {Gil-Mar{\'i}n}, H. and Gnedin, O. Y. and Gontcho, S. Gontcho A. and {Gonzalez-Morales}, A. X. and {Gonzalez-Perez}, V. and Gordon, C. and Graur, O. and Green, D. and Gruen, D. and Gsponer, R. and Guandalin, C. and Gutierrez, G. and Guy, J. and Hahn, C. and Han, J. J. and Han, J. and He, S. and {Herrera-Alcantar}, H. K. and Honscheid, K. and Hou, J. and Howlett, C. and Huterer, D. and Ir{\v s}i{\v c}, V. and Ishak, M. and Jacques, A. and Jimenez, J. and Jing, Y. P. and Joachimi, B. and Joudaki, S. and Joyce, R. and Jullo, E. and Juneau, S. and Kara{\c c}ayl{\i}, N. G. and Karim, T. and Kehoe, R. and Kent, S. and Khederlarian, A. and Kirkby, D. and Kisner, T. and Kitaura, F.-S. and Kizhuprakkat, N. and Kong, H. and Koposov, S. E. and Kremin, A. and Krolewski, A. and Lahav, O. and Lai, Y. and Lamman, C. and Lan, T.-W. and Landriau, M. and Lang, D. and Lange, J. U. and Lasker, J. and Goff, J. M. Le and Guillou, L. Le and Leauthaud, A. and Levi, M. E. and Li, S. and Li, T. S. and Lodha, K. and Lokken, M. and Luo, Y. and Magneville, C. and Manera, M. and Manser, C. J. and Margala, D. and Martini, P. and Maus, M. and McCullough, J. and McDonald, P. and Medina, G. E. and {Medina-Varela}, L. and Meisner, A. and {Mena-Fern{\'a}ndez}, J. and Menegas, A. and Mezcua, M. and Miquel, R. and {Montero-Camacho}, P. and Moon, J. and Moustakas, J. and {Mu{\~n}oz-Guti{\'e}rrez}, A. and {Mu{\~n}oz-Santos}, D. and Myers, A. D. and Myles, J. and Nadathur, S. and Najita, J. and Napolitano, L. and Newman, J. A. and Nikakhtar, F. and Nikutta, R. and Niz, G. and Noriega, H. E. and Padmanabhan, N. and Paillas, E. and {Palanque-Delabrouille}, N. and Palmese, A. and Pan, J. and Pan, Z. and Parkinson, D. and Peacock, J. and Percival, W. J. and {P{\'e}rez-Fern{\'a}ndez}, A. and {P{\'e}rez-R{\`a}fols}, I. and Peterson, P. and Piat, J. and Pieri, M. M. and Pinon, M. and Poppett, C. and Porredon, A. and Prada, F. and Pucha, R. and Qin, F. and Rabinowitz, D. and Raichoor, A. and {Ram{\'i}rez-P{\'e}rez}, C. and {Ramirez-Solano}, S. and Rashkovetskyi, M. and Ravoux, C. and Riley, A. H. and Rocher, A. and Rockosi, C. and Rohlf, J. and Ross, A. J. and Rossi, G. and Ruggeri, R. and {Ruhlmann-Kleider}, V. and Sabiu, C. G. and Said, K. and Saintonge, A. and Samushia, L. and Sanchez, E. and Sanders, N. and Saulder, C. and Schlafly, E. F. and Schlegel, D. and Scholte, D. and Schubnell, M. and Seo, H. and Shafieloo, A. and Sharples, R. and Silber, J. and Siudek, M. and Smith, A. and Sprayberry, D. and {Su{\'a}rez-P{\'e}rez}, J. and Swanson, J. and Tan, T. and Tarl{\'e}, G. and Taylor, P. and Thomas, G. and Tojeiro, R. and Turner, R. J. and Turner, W. and {Ure{\~n}a-L{\'o}pez}, L. A. and Vaisakh, R. and Valluri, M. and {Vargas-Maga{\~n}a}, M. and Verde, L. and Walther, M. and Wang, B. and Wang, M. S. and Wang, W. and Weaver, B. A. and Weaverdyck, N. and Wechsler, R. H. and White, M. and Wolfson, M. and Yang, J. and Y{\`e}che, C. and Youles, S. and Yu, J. and Yuan, S. and Zaborowski, E. A. and Zarrouk, P. and Zhang, H. and Zhao, C. and Zhao, R. and Zheng, Z. and Zhou, R. and Zou, H. and Zou, S. and Zu, Y.},
  year = {2025},
  month = mar,
  number = {arXiv:2503.14745},
  eprint = {2503.14745},
  primaryclass = {astro-ph},
  publisher = {arXiv},
  doi = {10.48550/arXiv.2503.14745},
  urldate = {2025-04-24},
  archiveprefix = {arXiv},
  langid = {american}
}

@article{durasUniversalBolometricCorrections2020,
  title = {Universal Bolometric Corrections for Active Galactic Nuclei over Seven Luminosity Decades},
  author = {Duras, F. and Bongiorno, A. and Ricci, F. and Piconcelli, E. and Shankar, F. and Lusso, E. and Bianchi, S. and Fiore, F. and Maiolino, R. and Marconi, A. and Onori, F. and Sani, E. and Schneider, R. and Vignali, C. and Franca, F. La},
  year = {2020},
  month = apr,
  journal = {Astronomy \& Astrophysics},
  volume = {636},
  pages = {A73},
  publisher = {EDP Sciences},
  issn = {0004-6361, 1432-0746},
  doi = {10.1051/0004-6361/201936817},
  urldate = {2025-08-01},
  copyright = {{\copyright} ESO 2020},
  langid = {english}
}

@article{edgarReviewBondiHoyle2004,
  title = {A Review of {{Bondi}}--{{Hoyle}}--{{Lyttleton}} Accretion},
  author = {Edgar, Richard},
  year = {2004},
  month = sep,
  journal = {New Astronomy Reviews},
  volume = {48},
  number = {10},
  pages = {843--859},
  issn = {1387-6473},
  doi = {10.1016/j.newar.2004.06.001},
  urldate = {2024-07-15}
}

@article{gerosaHierarchicalMergersStellarmass2021,
  title = {Hierarchical Mergers of Stellar-Mass Black Holes and Their Gravitational-Wave Signatures},
  author = {Gerosa, Davide and Fishbach, Maya},
  year = {2021},
  month = aug,
  journal = {Nature Astronomy},
  volume = {5},
  number = {8},
  pages = {749--760},
  publisher = {Nature Publishing Group},
  issn = {2397-3366},
  doi = {10.1038/s41550-021-01398-w},
  urldate = {2025-02-12},
  copyright = {2021 Springer Nature Limited},
  langid = {english}
}

@misc{gottliebSpinningGapDirectHorizon2025,
  title = {Spinning into the {{Gap}}: {{Direct-Horizon Collapse}} as the {{Origin}} of {{GW231123}} from {{End-to-End GRMHD Simulations}}},
  shorttitle = {Spinning into the {{Gap}}},
  author = {Gottlieb, Ore and Metzger, Brian D. and Issa, Danat and Li, Sean E. and Renzo, Mathieu and Isi, Maximiliano},
  year = {2025},
  month = aug,
  number = {arXiv:2508.15887},
  eprint = {2508.15887},
  primaryclass = {astro-ph},
  publisher = {arXiv},
  doi = {10.48550/arXiv.2508.15887},
  urldate = {2025-09-15},
  archiveprefix = {arXiv},
  langid = {american}
}

@article{grahamCandidateElectromagneticCounterpart2020,
  title = {Candidate {{Electromagnetic Counterpart}} to the {{Binary Black Hole Merger Gravitational-Wave Event S190521g}}},
  author = {Graham, M. J. and Ford, K. E. S. and McKernan, B. and Ross, N. P. and Stern, D. and Burdge, K. and Coughlin, M. and Djorgovski, S. G. and Drake, A. J. and Duev, D. and Kasliwal, M. and Mahabal, A. A. and {van Velzen}, S. and Belecki, J. and Bellm, E. C. and Burruss, R. and Cenko, S. B. and Cunningham, V. and Helou, G. and Kulkarni, S. R. and Masci, F. J. and Prince, T. and Reiley, D. and Rodriguez, H. and Rusholme, B. and Smith, R. M. and Soumagnac, M. T.},
  year = {2020},
  month = jun,
  journal = {Physical Review Letters},
  volume = {124},
  number = {25},
  pages = {251102},
  publisher = {American Physical Society},
  doi = {10.1103/PhysRevLett.124.251102},
  urldate = {2024-01-31},
  langid = {american}
}

@article{grahamLightDarkSearching2023,
  title = {A {{Light}} in the {{Dark}}: {{Searching}} for {{Electromagnetic Counterparts}} to {{Black Hole}}--{{Black Hole Mergers}} in {{LIGO}}/{{Virgo O3}} with the {{Zwicky Transient Facility}}},
  shorttitle = {A {{Light}} in the {{Dark}}},
  author = {Graham, Matthew J. and McKernan, Barry and Ford, K. E. Saavik and Stern, Daniel and Djorgovski, S. G. and Coughlin, Michael and Burdge, Kevin B. and Bellm, Eric C. and Helou, George and Mahabal, Ashish A. and Masci, Frank J. and Purdum, Josiah and Rosnet, Philippe and Rusholme, Ben},
  year = {2023},
  month = jan,
  journal = {The Astrophysical Journal},
  volume = {942},
  number = {2},
  pages = {99},
  publisher = {The American Astronomical Society},
  issn = {0004-637X},
  doi = {10.3847/1538-4357/aca480},
  urldate = {2025-04-19},
  langid = {english}
}

@article{guoPyQSOFitPythonCode2018,
  title = {{{PyQSOFit}}: {{Python}} Code to Fit the Spectrum of Quasars},
  shorttitle = {{{PyQSOFit}}},
  author = {Guo, Hengxiao and Shen, Yue and Wang, Shu},
  year = {2018},
  month = sep,
  journal = {Astrophysics Source Code Library},
  pages = {ascl:1809.008},
  urldate = {2025-09-04}
}

@article{harrisArrayProgrammingNumPy2020,
  title = {Array Programming with {{NumPy}}},
  author = {Harris, Charles R. and Millman, K. Jarrod and {van der Walt}, St{\'e}fan J. and Gommers, Ralf and Virtanen, Pauli and Cournapeau, David and Wieser, Eric and Taylor, Julian and Berg, Sebastian and Smith, Nathaniel J. and Kern, Robert and Picus, Matti and Hoyer, Stephan and {van Kerkwijk}, Marten H. and Brett, Matthew and Haldane, Allan and {del R{\'i}o}, Jaime Fern{\'a}ndez and Wiebe, Mark and Peterson, Pearu and {G{\'e}rard-Marchant}, Pierre and Sheppard, Kevin and Reddy, Tyler and Weckesser, Warren and Abbasi, Hameer and Gohlke, Christoph and Oliphant, Travis E.},
  year = {2020},
  month = sep,
  journal = {Nature},
  volume = {585},
  number = {7825},
  pages = {357--362},
  publisher = {Nature Publishing Group},
  issn = {1476-4687},
  doi = {10.1038/s41586-020-2649-2},
  urldate = {2025-04-18},
  copyright = {2020 The Author(s)},
  langid = {english}
}

@misc{heSystematicSearchAGN2025,
  title = {A {{Systematic Search}} for {{AGN Flares}} in {{ZTF Data Release}} 23},
  author = {He, Lei and Liu, Zheng-Yan and Niu, Rui and Zhou, Ming-Shen and Zou, Pu-Run and Gao, Bing-Zhou and Liang, Run-Duo and Zhu, Liang-Gui and Wang, Jian-Min and Jiang, Ning and Cai, Zhen-Yi and Jiang, Ji-an and Dai, Zi-Gao and Yuan, Ye-Fei and Chen, Yong-Jie and Zhao, Wen},
  year = {2025},
  month = jul,
  number = {arXiv:2507.20232},
  eprint = {2507.20232},
  primaryclass = {astro-ph},
  publisher = {arXiv, ApJS in press},
  doi = {10.48550/arXiv.2507.20232},
  urldate = {2025-07-29},
  archiveprefix = {arXiv},
  langid = {american}
}

@article{heTracingLightIdentification2025a,
  title = {Tracing the {{Light}}: {{Identification}} for the {{Optical Counterpart Candidates}} of {{Binary Black Holes}} during {{O3}}},
  shorttitle = {Tracing the {{Light}}},
  author = {He, Lei and Liu, Zhengyan and Niu, Rui and Gao, Bingzhou and Zhou, Mingshen and Zou, Purun and Liang, Runduo and Zhao, Wen and Jiang, Ning and Cai, Zhen-Yi and Dai, Zi-Gao and Yuan, Ye-Fei},
  year = {2025},
  month = sep,
  journal = {The Astrophysical Journal},
  volume = {990},
  number = {2},
  pages = {154},
  publisher = {The American Astronomical Society},
  issn = {0004-637X},
  doi = {10.3847/1538-4357/adf284},
  urldate = {2025-09-06},
  langid = {english}
}

@article{hoREVISEDCALIBRATIONVIRIAL2015,
  title = {A {{REVISED CALIBRATION OF THE VIRIAL MASS ESTIMATOR FOR BLACK HOLES IN ACTIVE GALAXIES BASED ON SINGLE-EPOCH H$\beta$ SPECTRA}}},
  author = {Ho, Luis C. and Kim, Minjin},
  year = {2015},
  month = aug,
  journal = {The Astrophysical Journal},
  volume = {809},
  number = {2},
  pages = {123},
  publisher = {The American Astronomical Society},
  issn = {0004-637X},
  doi = {10.1088/0004-637X/809/2/123},
  urldate = {2025-09-04},
  langid = {english}
}

@article{hunterMatplotlib2DGraphics2007,
  title = {Matplotlib: {{A 2D Graphics Environment}}},
  shorttitle = {Matplotlib},
  author = {Hunter, John D.},
  year = {2007},
  month = may,
  journal = {Computing in Science \& Engineering},
  volume = {9},
  number = {3},
  pages = {90--95},
  issn = {1558-366X},
  doi = {10.1109/MCSE.2007.55},
  urldate = {2025-04-18}
}

@article{ivezicLSSTScienceDrivers2019,
  title = {{{LSST}}: {{From Science Drivers}} to {{Reference Design}} and {{Anticipated Data Products}}},
  shorttitle = {{{LSST}}},
  author = {Ivezi{\'c}, {\v Z}eljko and Kahn, Steven M. and Tyson, J. Anthony and Abel, Bob and Acosta, Emily and Allsman, Robyn and Alonso, David and AlSayyad, Yusra and Anderson, Scott F. and Andrew, John and P. Angel, James Roger and Angeli, George Z. and Ansari, Reza and Antilogus, Pierre and Araujo, Constanza and Armstrong, Robert and Arndt, Kirk T. and Astier, Pierre and Aubourg, {\'E}ric and Auza, Nicole and Axelrod, Tim S. and Bard, Deborah J. and Barr, Jeff D. and Barrau, Aurelian and Bartlett, James G. and Bauer, Amanda E. and Bauman, Brian J. and Baumont, Sylvain and Bechtol, Ellen and Bechtol, Keith and Becker, Andrew C. and Becla, Jacek and Beldica, Cristina and Bellavia, Steve and Bianco, Federica B. and Biswas, Rahul and Blanc, Guillaume and Blazek, Jonathan and Blandford, Roger D. and Bloom, Josh S. and Bogart, Joanne and Bond, Tim W. and Booth, Michael T. and Borgland, Anders W. and Borne, Kirk and Bosch, James F. and Boutigny, Dominique and Brackett, Craig A. and Bradshaw, Andrew and Brandt, William Nielsen and Brown, Michael E. and Bullock, James S. and Burchat, Patricia and Burke, David L. and Cagnoli, Gianpietro and Calabrese, Daniel and Callahan, Shawn and Callen, Alice L. and Carlin, Jeffrey L. and Carlson, Erin L. and Chandrasekharan, Srinivasan and {Charles-Emerson}, Glenaver and Chesley, Steve and Cheu, Elliott C. and Chiang, Hsin-Fang and Chiang, James and Chirino, Carol and Chow, Derek and Ciardi, David R. and Claver, Charles F. and {Cohen-Tanugi}, Johann and Cockrum, Joseph J. and Coles, Rebecca and Connolly, Andrew J. and Cook, Kem H. and Cooray, Asantha and Covey, Kevin R. and Cribbs, Chris and Cui, Wei and Cutri, Roc and Daly, Philip N. and Daniel, Scott F. and Daruich, Felipe and Daubard, Guillaume and Daues, Greg and Dawson, William and Delgado, Francisco and Dellapenna, Alfred and de Peyster, Robert and de {Val-Borro}, Miguel and Digel, Seth W. and Doherty, Peter and Dubois, Richard and {Dubois-Felsmann}, Gregory P. and Durech, Josef and Economou, Frossie and Eifler, Tim and Eracleous, Michael and Emmons, Benjamin L. and Neto, Angelo Fausti and Ferguson, Henry and Figueroa, Enrique and {Fisher-Levine}, Merlin and Focke, Warren and Foss, Michael D. and Frank, James and Freemon, Michael D. and Gangler, Emmanuel and Gawiser, Eric and Geary, John C. and Gee, Perry and Geha, Marla and Gessner, Charles J. B. and Gibson, Robert R. and Gilmore, D. Kirk and Glanzman, Thomas and Glick, William and Goldina, Tatiana and Goldstein, Daniel A. and Goodenow, Iain and Graham, Melissa L. and Gressler, William J. and Gris, Philippe and Guy, Leanne P. and Guyonnet, Augustin and Haller, Gunther and Harris, Ron and Hascall, Patrick A. and Haupt, Justine and Hernandez, Fabio and Herrmann, Sven and Hileman, Edward and Hoblitt, Joshua and Hodgson, John A. and Hogan, Craig and Howard, James D. and Huang, Dajun and Huffer, Michael E. and Ingraham, Patrick and Innes, Walter R. and Jacoby, Suzanne H. and Jain, Bhuvnesh and Jammes, Fabrice and Jee, M. James and Jenness, Tim and Jernigan, Garrett and Jevremovi{\'c}, Darko and Johns, Kenneth and Johnson, Anthony S. and Johnson, Margaret W. G. and Jones, R. Lynne and {Juramy-Gilles}, Claire and Juri{\'c}, Mario and Kalirai, Jason S. and Kallivayalil, Nitya J. and Kalmbach, Bryce and Kantor, Jeffrey P. and Karst, Pierre and Kasliwal, Mansi M. and Kelly, Heather and Kessler, Richard and Kinnison, Veronica and Kirkby, David and Knox, Lloyd and Kotov, Ivan V. and Krabbendam, Victor L. and Krughoff, K. Simon and Kub{\'a}nek, Petr and Kuczewski, John and Kulkarni, Shri and Ku, John and Kurita, Nadine R. and Lage, Craig S. and Lambert, Ron and Lange, Travis and Langton, J. Brian and Guillou, Laurent Le and Levine, Deborah and Liang, Ming and Lim, Kian-Tat and Lintott, Chris J. and Long, Kevin E. and Lopez, Margaux and Lotz, Paul J. and Lupton, Robert H. and Lust, Nate B. and MacArthur, Lauren A. and Mahabal, Ashish and Mandelbaum, Rachel and Markiewicz, Thomas W. and Marsh, Darren S. and Marshall, Philip J. and Marshall, Stuart and May, Morgan and McKercher, Robert and McQueen, Michelle and Meyers, Joshua and Migliore, Myriam and Miller, Michelle and Mills, David J. and Miraval, Connor and Moeyens, Joachim and Moolekamp, Fred E. and Monet, David G. and Moniez, Marc and Monkewitz, Serge and Montgomery, Christopher and Morrison, Christopher B. and Mueller, Fritz and Muller, Gary P. and Arancibia, Freddy Mu{\~n}oz and Neill, Douglas R. and Newbry, Scott P. and Nief, Jean-Yves and Nomerotski, Andrei and Nordby, Martin and O'Connor, Paul and Oliver, John and Olivier, Scot S. and Olsen, Knut and O'Mullane, William and Ortiz, Sandra and Osier, Shawn and Owen, Russell E. and Pain, Reynald and Palecek, Paul E. and Parejko, John K. and Parsons, James B. and Pease, Nathan M. and Peterson, J. Matt and Peterson, John R. and Petravick, Donald L. and Petrick, M. E. Libby and Petry, Cathy E. and Pierfederici, Francesco and Pietrowicz, Stephen and Pike, Rob and Pinto, Philip A. and Plante, Raymond and Plate, Stephen and Plutchak, Joel P. and Price, Paul A. and Prouza, Michael and Radeka, Veljko and Rajagopal, Jayadev and Rasmussen, Andrew P. and Regnault, Nicolas and Reil, Kevin A. and Reiss, David J. and Reuter, Michael A. and Ridgway, Stephen T. and Riot, Vincent J. and Ritz, Steve and Robinson, Sean and Roby, William and Roodman, Aaron and Rosing, Wayne and Roucelle, Cecille and Rumore, Matthew R. and Russo, Stefano and Saha, Abhijit and Sassolas, Benoit and Schalk, Terry L. and Schellart, Pim and Schindler, Rafe H. and Schmidt, Samuel and Schneider, Donald P. and Schneider, Michael D. and Schoening, William and Schumacher, German and Schwamb, Megan E. and Sebag, Jacques and Selvy, Brian and Sembroski, Glenn H. and Seppala, Lynn G. and Serio, Andrew and Serrano, Eduardo and Shaw, Richard A. and Shipsey, Ian and Sick, Jonathan and Silvestri, Nicole and Slater, Colin T. and Smith, J. Allyn and Smith, R. Chris and Sobhani, Shahram and Soldahl, Christine and {Storrie-Lombardi}, Lisa and Stover, Edward and Strauss, Michael A. and Street, Rachel A. and Stubbs, Christopher W. and Sullivan, Ian S. and Sweeney, Donald and Swinbank, John D. and Szalay, Alexander and Takacs, Peter and Tether, Stephen A. and Thaler, Jon J. and Thayer, John Gregg and Thomas, Sandrine and Thornton, Adam J. and Thukral, Vaikunth and Tice, Jeffrey and Trilling, David E. and Turri, Max and Berg, Richard Van and Berk, Daniel Vanden and Vetter, Kurt and Virieux, Francoise and Vucina, Tomislav and Wahl, William and Walkowicz, Lucianne and Walsh, Brian and Walter, Christopher W. and Wang, Daniel L. and Wang, Shin-Yawn and Warner, Michael and Wiecha, Oliver and Willman, Beth and Winters, Scott E. and Wittman, David and Wolff, Sidney C. and {Wood-Vasey}, W. Michael and Wu, Xiuqin and Xin, Bo and Yoachim, Peter and Zhan, Hu},
  year = {2019},
  month = mar,
  journal = {The Astrophysical Journal},
  volume = {873},
  number = {2},
  pages = {111},
  publisher = {The American Astronomical Society},
  issn = {0004-637X},
  doi = {10.3847/1538-4357/ab042c},
  urldate = {2025-06-20},
  langid = {english}
}

@article{karasEnhancedActivityMassive2007,
  title = {Enhanced Activity of Massive Black Holes by Stellar Capture Assisted by a Self-Gravitating Accretion Disc},
  author = {Karas, V. and {\v S}ubr, L.},
  year = {2007},
  month = jul,
  journal = {Astronomy \& Astrophysics},
  volume = {470},
  number = {1},
  pages = {11--19},
  publisher = {EDP Sciences},
  issn = {0004-6361, 1432-0746},
  doi = {10.1051/0004-6361:20066068},
  urldate = {2025-07-16},
  copyright = {{\copyright} ESO, 2007},
  langid = {english}
}

@article{kassBayesFactors1995,
  title = {Bayes {{Factors}}},
  author = {Kass, Robert E. and Raftery, Adrian E.},
  year = {1995},
  month = jun,
  journal = {Journal of the American Statistical Association},
  volume = {90},
  number = {430},
  pages = {773--795},
  issn = {0162-1459, 1537-274X},
  doi = {10.1080/01621459.1995.10476572},
  urldate = {2025-01-15},
  langid = {english}
}

@article{liCorecollapseSupernovaExplosions2023,
  title = {Core-Collapse {{Supernova Explosions}} in {{Active Galactic Nucleus Accretion Disks}}},
  author = {Li, Fu-Lin and Liu, Yu and Fan, Xiao and Hu, Mao-Kai and Yang, Xuan and Geng, Jin-Jun and Wu, Xue-Feng},
  year = {2023},
  month = jun,
  journal = {The Astrophysical Journal},
  volume = {950},
  number = {2},
  pages = {161},
  publisher = {The American Astronomical Society},
  issn = {0004-637X},
  doi = {10.3847/1538-4357/acd2d1},
  urldate = {2025-04-16},
  langid = {english}
}

@misc{liGW231123ProductSuccessive2025,
  title = {{{GW231123}}: A Product of Successive Mergers from \${\textbackslash}sim 10 \$ Stellar-Mass Black Holes},
  shorttitle = {{{GW231123}}},
  author = {Li, Yin-Jie and Tang, Shan-Peng and Xue, Ling-Qin and Fan, Yi-Zhong},
  year = {2025},
  month = jul,
  number = {arXiv:2507.17551},
  eprint = {2507.17551},
  primaryclass = {astro-ph},
  publisher = {arXiv},
  doi = {10.48550/arXiv.2507.17551},
  urldate = {2025-07-24},
  archiveprefix = {arXiv},
  langid = {american}
}

@misc{liHierarchicalMergerScenario2025,
  title = {The {{Hierarchical Merger Scenario}} for {{GW231123}}},
  author = {Li, Guo-Peng and Fan, Xi-Long},
  year = {2025},
  month = sep,
  number = {arXiv:2509.08298},
  eprint = {2509.08298},
  primaryclass = {astro-ph},
  publisher = {arXiv},
  doi = {10.48550/arXiv.2509.08298},
  urldate = {2025-09-13},
  archiveprefix = {arXiv},
  langid = {american}
}

@misc{liuFormationGW231123Population2025,
  title = {On the {{Formation}} of {{GW231123}} in {{Population III Star Clusters}}},
  author = {Liu, Shuai and Wang, Long and Tanikawa, Ataru and Wu, Weiwei and Fujii, Michiko S.},
  year = {2025},
  month = oct,
  number = {arXiv:2510.05634},
  eprint = {2510.05634},
  primaryclass = {astro-ph},
  publisher = {arXiv},
  doi = {10.48550/arXiv.2510.05634},
  urldate = {2025-10-08},
  archiveprefix = {arXiv},
  langid = {american}
}

@article{liUseBinaryBlack2025,
  title = {The {{Use}} of {{Binary Black Holes Merging}} in {{Active Galactic Nuclei Disks}} for {{Hubble Constant Measurements}}},
  author = {Li, Guo-Peng and Fan, Xi-Long},
  year = {2025},
  month = jun,
  journal = {The Astrophysical Journal},
  volume = {986},
  number = {1},
  pages = {61},
  publisher = {The American Astronomical Society},
  issn = {0004-637X},
  doi = {10.3847/1538-4357/adda46},
  urldate = {2025-08-12},
  langid = {english}
}

@misc{lucaGW231123PossiblePrimordial2025,
  title = {{{GW231123}}: A {{Possible Primordial Black Hole Origin}}},
  shorttitle = {{{GW231123}}},
  author = {Luca, Valerio De and Franciolini, Gabriele and Riotto, Antonio},
  year = {2025},
  month = aug,
  number = {arXiv:2508.09965},
  eprint = {2508.09965},
  primaryclass = {astro-ph},
  publisher = {arXiv},
  doi = {10.48550/arXiv.2508.09965},
  urldate = {2025-08-18},
  archiveprefix = {arXiv},
  langid = {american}
}

@incollection{mapelliFormationChannelsSingle2021,
  title = {Formation Channels of Single and Binary Stellar-Mass Black Holes},
  booktitle = {Handbook of {{Gravitational Wave Astronomy}}},
  author = {Mapelli, Michela},
  year = {2021},
  eprint = {2106.00699},
  primaryclass = {astro-ph},
  pages = {1--65},
  publisher = {Springer Singapore},
  doi = {10.1007/978-981-15-4702-7_16-1},
  urldate = {2025-01-17},
  archiveprefix = {arXiv},
  langid = {american}
}

@misc{masciNewForcedPhotometry2023,
  title = {A {{New Forced Photometry Service}} for the {{Zwicky Transient Facility}}},
  author = {Masci, Frank J. and Laher, Russ R. and Rusholme, Benjamin and Shupe, David and Paladini, Roberta and Groom, Steve and Wold, Avery and Miller, Adam A. and Drake, Andrew},
  year = {2023},
  month = jul,
  number = {arXiv:2305.16279},
  eprint = {2305.16279},
  primaryclass = {astro-ph},
  publisher = {arXiv},
  urldate = {2024-09-11},
  archiveprefix = {arXiv},
  langid = {american}
}

@article{mckernanRampressureStrippingKicked2019,
  title = {Ram-Pressure {{Stripping}} of a {{Kicked Hill Sphere}}: {{Prompt Electromagnetic Emission}} from the {{Merger}} of {{Stellar Mass Black Holes}} in an {{AGN Accretion Disk}}},
  shorttitle = {Ram-Pressure {{Stripping}} of a {{Kicked Hill Sphere}}},
  author = {McKernan, B. and Ford, K. E. S. and Bartos, I. and Graham, M. J. and Lyra, W. and Marka, S. and Marka, Z. and Ross, N. P. and Stern, D. and Yang, Y.},
  year = {2019},
  month = oct,
  journal = {The Astrophysical Journal Letters},
  volume = {884},
  number = {2},
  pages = {L50},
  publisher = {The American Astronomical Society},
  issn = {2041-8205},
  doi = {10.3847/2041-8213/ab4886},
  urldate = {2024-01-31},
  langid = {english}
}

@article{mckinneyDataStructuresStatistical2010,
  title = {Data {{Structures}} for {{Statistical Computing}} in {{Python}}},
  author = {McKinney, Wes},
  year = {2010},
  month = may,
  journal = {scipy},
  doi = {10.25080/Majora-92bf1922-00a},
  urldate = {2025-07-18},
  langid = {english}
}

@article{mortonGW190521BinaryBlack2023,
  title = {{{GW190521}}: {{A}} Binary Black Hole Merger inside an Active Galactic Nucleus?},
  shorttitle = {{{GW190521}}},
  author = {Morton, Sophia L. and Rinaldi, Stefano and {Torres-Orjuela}, Alejandro and Derdzinski, Andrea and Vaccaro, M. Paola and Del Pozzo, Walter},
  year = {2023},
  month = dec,
  journal = {Physical Review D},
  volume = {108},
  number = {12},
  pages = {123039},
  publisher = {American Physical Society},
  doi = {10.1103/PhysRevD.108.123039},
  urldate = {2025-01-08},
  langid = {american}
}

@misc{mukherjeeFirstMeasurementHubble2020,
  title = {First Measurement of the {{Hubble}} Parameter from Bright Binary Black Hole {{GW190521}}},
  author = {Mukherjee, Suvodip and Ghosh, Archisman and Graham, Matthew J. and Karathanasis, Christos and Kasliwal, Mansi M. and Maga{\~n}a Hernandez, Ignacio and Nissanke, Samaya M. and Silvestri, Alessandra and Wandelt, Benjamin D.},
  year = {2020},
  month = sep,
  journal = {arXiv e-prints},
  doi = {10.48550/arXiv.2009.14199},
  urldate = {2025-01-16}
}

@misc{nojiriDynamicalBlackHole2025,
  title = {Dynamical {{Black Hole}} in the Accelerating {{Universe}} Approaching the Future Singularity -- {{Possible}} Origin of (Super-)Massive Black Holes},
  author = {Nojiri, Shin'ichi and Odintsov, Sergei D.},
  year = {2025},
  month = sep,
  number = {arXiv:2508.07524},
  eprint = {2508.07524},
  primaryclass = {gr-qc},
  publisher = {arXiv},
  doi = {10.48550/arXiv.2508.07524},
  urldate = {2025-09-26},
  archiveprefix = {arXiv},
  langid = {american}
}

@article{palmeseLIGOVirgoBlack2021,
  title = {Do {{LIGO}}/{{Virgo}} Black Hole Mergers Produce {{AGN}} Flares? {{The}} Case of {{GW190521}} and Prospects for Reaching a Confident Association},
  shorttitle = {Do {{LIGO}}/{{Virgo}} Black Hole Mergers Produce {{AGN}} Flares?},
  author = {Palmese, Antonella and Fishbach, Maya and Burke, Colin J. and Annis, James T. and Liu, Xin},
  year = {2021},
  month = jun,
  journal = {The Astrophysical Journal Letters},
  volume = {914},
  number = {2},
  eprint = {2103.16069},
  primaryclass = {astro-ph},
  pages = {L34},
  issn = {2041-8205, 2041-8213},
  doi = {10.3847/2041-8213/ac0883},
  urldate = {2024-04-12},
  archiveprefix = {arXiv}
}

@misc{popaVeryMassiveRapidly2025,
  title = {Very {{Massive}}, {{Rapidly Spinning Binary Black Hole Progenitors}} through {{Chemically Homogeneous Evolution}} -- {{The Case}} of {{GW231123}}},
  author = {Popa, Silvia A. and de Mink, Selma E.},
  year = {2025},
  month = aug,
  number = {arXiv:2509.00154},
  eprint = {2509.00154},
  primaryclass = {astro-ph},
  publisher = {arXiv},
  doi = {10.48550/arXiv.2509.00154},
  urldate = {2025-09-03},
  archiveprefix = {arXiv},
  langid = {american}
}

@article{ryuInplaneTidalDisruption2024,
  title = {In-Plane Tidal Disruption of Stars in Discs of Active Galactic Nuclei},
  author = {Ryu, Taeho and McKernan, Barry and Ford, K E Saavik and Cantiello, Matteo and Graham, Matthew and Stern, Daniel and Leigh, Nathan W C},
  year = {2024},
  month = jan,
  journal = {Monthly Notices of the Royal Astronomical Society},
  volume = {527},
  number = {3},
  pages = {8103--8117},
  issn = {0035-8711},
  doi = {10.1093/mnras/stad3487},
  urldate = {2025-07-06},
  langid = {american}
}

@article{sanchez-saezAlertClassificationALeRCE2021,
  title = {Alert {{Classification}} for the {{ALeRCE Broker System}}: {{The Light Curve Classifier}}},
  shorttitle = {Alert {{Classification}} for the {{ALeRCE Broker System}}},
  author = {{S{\'a}nchez-S{\'a}ez}, P. and Reyes, I. and Valenzuela, C. and F{\"o}rster, F. and Eyheramendy, S. and Elorrieta, F. and Bauer, F. E. and {Cabrera-Vives}, G. and Est{\'e}vez, P. A. and Catelan, M. and Pignata, G. and Huijse, P. and De Cicco, D. and Ar{\'e}valo, P. and {Carrasco-Davis}, R. and Abril, J. and Kurtev, R. and Borissova, J. and Arredondo, J. and {Castillo-Navarrete}, E. and Rodriguez, D. and {Ruz-Mieres}, D. and Moya, A. and {Sabatini-Gacit{\'u}a}, L. and {Sep{\'u}lveda-Cobo}, C. and {Camacho-I{\~n}iguez}, E.},
  year = {2021},
  month = mar,
  journal = {The Astronomical Journal},
  volume = {161},
  pages = {141},
  publisher = {IOP},
  issn = {0004-6256},
  doi = {10.3847/1538-3881/abd5c1},
  urldate = {2025-04-13},
  langid = {american}
}

@misc{sdsscollaborationNineteenthDataRelease2025,
  title = {The {{Nineteenth Data Release}} of the {{Sloan Digital Sky Survey}}},
  author = {{SDSS Collaboration} and Pallathadka, Gautham Adamane and Aghakhanloo, Mojgan and Aird, James and Almeida, Andr{\'e}s and Amrita, Singh and Anders, Friedrich and Anderson, Scott F. and Arseneau, Stefan and Avila, Consuelo Gonz{\'a}lez and Aviram, Shir and Aydar, Catarina and Badenes, Carles and {Barrera-Ballesteros}, Jorge K. and Bauer, Franz E. and Behmard, Aida and Berg, Michelle and Besser, F. and Bidin, Christian Moni and Bizyaev, Dmitry and Blanc, Guillermo and Blanton, Michael R. and Bovy, Jo and Brandt, William Nielsen and Brownstein, Joel R. and Buchner, Johannes and Bulbul, Esra and Burchett, Joseph N. and Carigi, Leticia and Carlberg, Joleen K. and Casey, Andrew R. and Chakraborty, Priyanka and Chanam{\'e}, Julio and Chandra, Vedant and Chiappini, Cristina and Chilingarian, Igor and Comparat, Johan and Covey, Kevin and Crumpler, Nicole and Cunha, Katia and D'Onghia, Elena and Dai, Xinyu and Darling, Jeremy and Davis, Megan and Lee, Nathan De and Deacon, Niall and Delgado, Jos{\'e} Eduardo M{\'e}ndez and Demasi, Sebastian and Demianenko, Mariia and Demke, Delvin and Donor, John and Drory, Niv and Durango, Monica Alejandra Villa and Dwelly, Tom and Egorov, Oleg and Egorova, Evgeniya and {El-Badry}, Kareem and Eracleous, Mike and Fan, Xiaohui and Farr, Emily and Finkbeiner, Douglas P. and Fries, Logan and Frinchaboy, Peter and Fusillo, Nicola Pietro Gentile and F{\'e}lix, Luis Daniel Serrano and Gaensicke, Boris and Galligan, Emma and Garc{\'i}a, Pablo and Gelfand, Joseph and Grabowski, Katie and Grebel, Eva and Green, Paul J. and Greve, Hannah and Grier, Catherine and Griffith, Emily and Guetzoyan, Paloma and Gupta, Pramod and Hackshaw, Zoe and Hall, Patrick B. and Hawkins, Keith and Heged{\H u}s, Viola and Hekker, Saskia and Herbst, T. M. and Hermes, J. J. and {Hern{\'a}ndez-Garc{\'i}a}, Lorena and Hiremath, Pranavi and Hogg, David W. and Holtzman, Jon and Horne, Keith and Horta, Danny and Huang, Yang and Hutchinson, Brian and H{\"a}berle, Maximilian and {Ibarra-Medel}, Hector Javier and Ji, Alexander P. and Jofre, Paula and Johnson, James W. and Johnson, Jennifer and Johnston, Evelyn J. and Kaldor, Mary and Katkov, Ivan and Khalatyan, Arman and Khoperskov, Sergey and Klessen, Ralf and Kluge, Matthias and Koekemoer, Anton M. and Kollmeier, Juna A. and Kounkel, Marina and Kreckel, Kathryn and Krishnarao, Dhanesh and Krumpe, Mirko and Lacerna, Ivan and Laporte, Chervin and Lepine, Sebastien and Li, Jing and Liang, Fu-Heng and Limberg, Guilherme and Liu, Xin and Loebman, Sarah and Long, Knox and Lu, Yuxi and Lucey, Madeline and {Lugo-Aranda}, Alejandra Z. and {Martinez-Aldama}, Mary Loli Mart{\'i}nez and McKinnon, Kevin and Medan, Ilija and Merloni, Andrea and Morrison, Sean and Myers, Natalie and M{\'e}sz{\'a}ros, Szabolcs and {M{\"u}ller-Horn}, Johanna and Nepal, Samir and Ness, Melissa and Nidever, David and Nitschelm, Christian and Oravetz, Audrey and Otto, Jonah and Pan, Kaike and Paolino, Facundo P{\'e}rez and Pe{\~n}aloza, Castalia Alenka Negrete and Pinsonneault, Marc and Popp, Manuchehr Taghizadeh and {Price-Whelan}, Adrian and Pulatova, Nadiia and Queiroz, Anna Barbara and Raddick, Jordan and Rankine, Amy and Rix, Hans-Walter and {Rom{\'a}n-Z{\'u}{\~n}iga}, Carlos and Rosso, Daniela Fern{\'a}ndez and Runnoe, Jessie and Saad, Serat Mahmud and Salvato, Mara and Sanchez, Sebastian F. and Sattler, Natascha and Saydjari, Andrew and Sayres, Conor and Schlaufman, Kevin and Schneider, Donald P. and Schwope, Axel and Seaton, Lucas M. and Seeburger, Rhys and Serna, Javier and Sharma, Sanjib and Shen, Yue and Sinha, Amaya and Sizemore, Brian and Sniegowska, Marzena and Song, Yingyi and Souto, Diogo and Stassun, Keivan and Steinmetz, Matthias and Stone, Zachary and {Stone-Martinez}, Alexander and Stringfellow, Guy S. and S{\'a}nchez, Aurora Mata and {S{\'a}nchez-Gallego}, Jos{\'e} and Tan, Jonathan and Tayar, Jamie and Thai, Riley and Thakar, Ani and Thibodeaux, Pierre and Ting, Yuan-Sen and Tkachenko, Andrew and Trakhtenbrot, Benny and Trincado, Jose G. Fernandez and Troup, Nicholas and Trump, Jonathan R. and Ulloa, Natalie and der Marel, Roeland P. Van and Vera, Pablo and Villanova, Sandro and Villase{\~n}or, Jaime and Wang, Ji and Way, Zachary and Weijmans, Anne-Marie and Wheeler, Adam and Wilson, John C. and Wofford, Aida and Wong, Tony and Wu, Qiaoya and Wylezalek, Dominika and Xue, Xiang-Xiang and Yan, Renbin and Yang, Qian and Zakamska, Nadia and Zari, Eleonora and Zasowski, Gail and Zeltyn, Grisha and Zheng, Zheng and Zucker, Catherine and Zerme{\~n}o, Rodolfo de J.},
  year = {2025},
  month = jul,
  number = {arXiv:2507.07093},
  eprint = {2507.07093},
  primaryclass = {astro-ph},
  publisher = {arXiv},
  doi = {10.48550/arXiv.2507.07093},
  urldate = {2025-09-23},
  archiveprefix = {arXiv}
}

@misc{shanDegeneracyMicrolensingWave2025,
  title = {The Degeneracy between Microlensing Wave Effect and Precession in Strongly Lensed Gravitational Wave},
  author = {Shan, Xikai and Yang, Huan and Mao, Shude and Hannuksela, Otto A.},
  year = {2025},
  month = sep,
  number = {arXiv:2508.21262},
  eprint = {2508.21262},
  primaryclass = {astro-ph},
  publisher = {arXiv},
  doi = {10.48550/arXiv.2508.21262},
  urldate = {2025-09-26},
  archiveprefix = {arXiv},
  langid = {american}
}

@article{singerRapidBayesianPosition2016,
  title = {Rapid {{Bayesian}} Position Reconstruction for Gravitational-Wave Transients},
  author = {Singer, Leo P. and Price, Larry R.},
  year = {2016},
  month = jan,
  journal = {Physical Review D},
  volume = {93},
  number = {2},
  pages = {024013},
  publisher = {American Physical Society},
  doi = {10.1103/PhysRevD.93.024013},
  urldate = {2025-02-12},
  langid = {american}
}

@misc{stegmannResolvingBlackHole2025,
  title = {Resolving {{Black Hole Family Issues Among}} the {{Massive Ancestors}} of {{Very High-Spin Gravitational-Wave Events Like GW231123}}},
  author = {Stegmann, Jakob and Olejak, Aleksandra and de Mink, Selma E.},
  year = {2025},
  month = jul,
  number = {arXiv:2507.15967},
  eprint = {2507.15967},
  primaryclass = {astro-ph},
  publisher = {arXiv},
  doi = {10.48550/arXiv.2507.15967},
  urldate = {2025-07-23},
  archiveprefix = {arXiv},
  langid = {american}
}

@article{tagawaShockCoolingBreakout2024,
  title = {Shock {{Cooling}} and {{Breakout Emission}} for {{Optical Flares Associated}} with {{Gravitational-wave Events}}},
  author = {Tagawa, Hiromichi and Kimura, Shigeo S and Haiman, Zolt{\'a}n and Perna, Rosalba and Bartos, Imre},
  year = {2024},
  month = apr,
  journal = {The Astrophysical Journal},
  volume = {966},
  number = {1},
  pages = {21},
  publisher = {The American Astronomical Society},
  issn = {0004-637X},
  doi = {10.3847/1538-4357/ad2e0b},
  urldate = {2025-01-17},
  langid = {english}
}

@misc{tanikawaGW231123FormationPopulation2025,
  title = {{{GW231123 Formation}} from {{Population III Stars}}: {{Isolated Binary Evolution}}},
  shorttitle = {{{GW231123 Formation}} from {{Population III Stars}}},
  author = {Tanikawa, Ataru and Liu, Shuai and Wu, WeiWei and Fujii, Michiko S. and Wang, Long},
  year = {2025},
  month = aug,
  number = {arXiv:2508.01135},
  eprint = {2508.01135},
  primaryclass = {astro-ph},
  publisher = {arXiv},
  doi = {10.48550/arXiv.2508.01135},
  urldate = {2025-09-15},
  archiveprefix = {arXiv},
  langid = {american}
}

@misc{theligoscientificcollaborationGW231123BinaryBlack2025,
  title = {{{GW231123}}: A {{Binary Black Hole Merger}} with {{Total Mass}} 190-265 \${{M}}\_\{{\textbackslash}odot\}\$},
  shorttitle = {{{GW231123}}},
  author = {{The LIGO Scientific Collaboration} and {The Virgo Collaboration} and {The KAGRA Collaboration}},
  year = {2025},
  month = jul,
  number = {arXiv:2507.08219},
  eprint = {2507.08219},
  primaryclass = {astro-ph},
  publisher = {arXiv},
  doi = {10.48550/arXiv.2507.08219},
  urldate = {2025-07-14},
  archiveprefix = {arXiv},
  langid = {american}
}

@incollection{thompsonFermiGammaRaySpace2023,
  title = {Fermi {{Gamma-Ray Space Telescope}}},
  booktitle = {Handbook of {{X-ray}} and {{Gamma-ray Astrophysics}}},
  author = {Thompson, David J. and {Wilson-Hodge}, Colleen A.},
  year = {2023},
  pages = {1--31},
  publisher = {Springer, Singapore},
  doi = {10.1007/978-981-16-4544-0_58-1},
  urldate = {2025-10-13},
  isbn = {978-981-16-4544-0},
  langid = {english}
}

@article{vajpeyiMeasuringPropertiesActive2022,
  title = {Measuring the {{Properties}} of {{Active Galactic Nuclei Disks}} with {{Gravitational Waves}}},
  author = {Vajpeyi, Avi and Thrane, Eric and Smith, Rory and McKernan, Barry and Saavik Ford, K. E.},
  year = {2022},
  month = may,
  journal = {The Astrophysical Journal},
  volume = {931},
  number = {2},
  pages = {82},
  publisher = {The American Astronomical Society},
  issn = {0004-637X},
  doi = {10.3847/1538-4357/ac6180},
  urldate = {2025-03-08},
  langid = {english}
}
\bibliographystyle{aasjournalv7}
\end{document}